\documentclass[usenatbib, twocolumn, twocolappendix]{aastex631}

\usepackage{graphicx}	
\usepackage{amsmath}	
\usepackage{ulem}
\usepackage{soul}
\usepackage[flushleft]{threeparttable}

\hypersetup{linkcolor=red, citecolor=blue, filecolor=cyan, urlcolor=magenta}

\newcommand{\rmd}{{\mathrm d}}
\newcommand{\deltak}{\delta_\mathbf{k}}
\newcommand{\phik}{\phi_\mathbf{k}}
\newcommand{\vect}[1]{\mathbf{#1}}
\newcommand{\avg}[1]{\langle{#1}\rangle}
\newcommand{\abs}[1]{\left\vert{#1}\right\vert}

\newcommand{\kvec}{{\mathbf{k}}}
\newcommand{\lhmpc}{{h^{-1}\rm Mpc}}
\newcommand{\ihmpc}{{h\ \rm Mpc^{-1}}}
\newcommand{\lhkpc}{{h^{-1}\rm kpc}}
\newcommand{\msun}{{h^{-1}\rm M_{\odot}}}

\shorttitle{One-point Statistics of the Ratio of Two Fourier-transformed Cosmic Fields}
\shortauthors{Li et al.}

\graphicspath{{./}{figures/}}

\begin{document}

\title{About One-point Statistics of the Ratio of Two Fourier-transformed Cosmic Fields and an Application}

\correspondingauthor{Ming Li, Jun Pan}
\email{mingli@nao.cas.cn, jpan@bao.ac.cn}

\author{Ming Li} 
\affiliation{National Astronomical Observatories, Chinese Academy of Sciences, Beijing 100101, People's Republic of China}

\author{Jun Pan}
\affiliation{National Astronomical Observatories, Chinese Academy of Sciences, Beijing 100101, People's Republic of China}

\author{Pengjie Zhang}
\affiliation{Department of Astronomy, School of Physics and Astronomy, Shanghai Jiao Tong University, Shanghai  200240, People's Republic of China}
\affiliation{Key Laboratory for Particle Astrophysics and Cosmology (MOE)/Shanghai Key Laboratory for Particle Physics and Cosmology, People's Republic of China}
\affiliation{Tsung-Dao Lee Institute, Shanghai Jiao Tong University, Shanghai, 200240, People's Republic of China}

\author{Jie Wang}
\affiliation{National Astronomical Observatories, Chinese Academy of Sciences, Beijing 100101, People's Republic of China}

\author{Longlong Feng}
\affiliation{School of Physics and Astronomy, Sun Yat-Sen University, Zhuhai Campus, No. 2, Daxue Road Zhuhai, Guangdong, 519082, People's Republic of China}

\author{Liang Gao}
\affiliation{National Astronomical Observatories, Chinese Academy of Sciences, Beijing 100101, People's Republic of China}

\author{Xi Kang}
\affiliation{Zhejiang University-Purple Mountain Observatory Joint Research Center for Astronomy, Zhejiang University, Hangzhou 310027, People's Republic of China}
\affiliation{Purple Mountain Observatory, CAS, No.8 Yuanhua Road, Qixia District, Nanjing 210034, People's Republic of China}

\author{Guoliang Li} 
\affiliation{Purple Mountain Observatory, CAS, No.8 Yuanhua Road, Qixia District, Nanjing 210034, People's Republic of China}
\affiliation{National Basic Science Data Center, Building No.2, 4, Zhongguancun South 4th Street, Beijing 100190, People's Republic of China}

\author{Weipeng Lin}
\affiliation{School of Physics and Astronomy, Sun Yat-Sen University, Zhuhai Campus, No. 2, Daxue Road Zhuhai, Guangdong, 519082, People's Republic of China}
 
\author{Haihui Wang}
\affiliation{School of Mathematical Sciences, Beihang University, 37 Xueyuan Road, Beijing 100191, People's Republic of China}

\begin{abstract}
The Fourier transformation is an effective and efficient operation of Gaussianization at the one-point level. Using a set of N-body simulation data, we verified that the one-point distribution functions of the dark matter momentum divergence and density fields closely follow complex Gaussian distributions. The one-point distribution function of the quotient of two complex Gaussian variables is introduced and studied. Statistical theories are then applied to model one-point statistics about the growth of individual Fourier mode of the dark matter density field, which can be obtained by the ratio of two Fourier transformed cosmic fields. Our simulation results proved that the models based on the Gaussian approximation are impressively accurate, and our analysis revealed many interesting aspects about the growth of dark matter's density fluctuation in Fourier space. 
\end{abstract}



\section{Introduction}
\label{sec:intro}
Understanding the statistical properties of the inhomogeneity in the spatial distribution of dark matter and its evolution in an expanding universe is one of the most crucial subjects in cosmology. The growing process of density fluctuation is not simple, even for a system composed of collisionless dark matter particles solely. Complexities lie in many aspects. Statistically, though the primordial density fluctuation is assumed to be Gaussian or nearly so with possibly a small amount of primordial non-Gaussianity \citep[see the review of ][and references therein]{Chen2010}, the nonlinear gravitation evolution and other physical processes would drive the distribution of late-time dark matter to be highly non-Gaussian. Gaussian distribution is mathematically simple for practical application and can be fully described by its first two moments. However, the non-Gaussianity means that an entire hierarchy of higher-order cumulants or correlation functions is awaiting exploration. Furthermore, it has been shown that the distribution of dark matter is very close to the lognormal distribution and cannot be completely specified by its moments \citep{ColesJones1991, Carron2011}. 

Non-Gaussianity makes measurement complex and difficult. If one intends to extract information from late-time density fields, statistics beyond two-point level \citep[][]{ScoccimarroEtal2001,BernardeauEtal2002,SefusattiEtal2006} are needed. In attempts to simplify the analysis, several Gaussianization schemes have been proposed and applied with notable success \citep[e.g.][]{NeyrinckEtal2009, ScherrerEtal2010, YuEtal2011, CarronSzapudi2013}. 
The schemes really help reduce the non-Gaussianity and enhance the cosmological information in the two-point statistics \citep[e.g.][]{YuEtal2016,ReppSzapudi2018}. The essence of Gaussianization is to perform a local transformation, so that the non-Gaussianity in statistics of the transformed field at the first few low orders might be significantly suppressed. Then the Gaussian approximation could be applied to model the corresponding statistics. One has to keep in mind that non-Gaussianity never vanishes, but instead has a different appearance \citep{QinEtal2020}.

\citet{Matsubara2007} formally derived that, for an arbitrary random field in a spatially homogeneous and sufficiently large space, the one-point probability distribution function (one-point PDF) of its Fourier mode shall approach Gaussian, provided that the polyspectra $P^{(2n)}(\vect{k}, \ldots, \vect{k}; -\vect{k}, \ldots, -\vect{k})$ are finite for any positive integer $n$. For a cosmic density field, spatial homogeneity is normally ensured. The condition of finite polyspectra at arbitrary even orders has not been exhaustively verified. But non-Gaussianity in the density field on most scales with $\vect{k}\neq 0$ is believed to be finite, since the volume average correlation functions derived from cosmological simulations and galaxy surveys are always finite up to detected ranks \citep[e.g.][]{MeiksinEtal1992, BouchetHernquist1992, GaztanagaFrieman1994, CrotonEtal2004,  HellwingEtal2010, CappiEtal2015}. Actually, part of the conclusions of \citet{Matsubara2007} have already been proved to be applicable to the cosmic density field in N-body simulations by \citet{HikageEtal2004}.

\citet{FalckEtal2021} recently measured the one-point PDFs of Fourier modes of the dark matter density fields of the {\tt Indra} simulation suite. They found that the Gaussian approximation of \citet{Matsubara2007} is consistent with simulations on scales as large as $\abs{\vect{k}}\approx 0.6\ \ihmpc$ with relatively good accuracy. Another independent research of \citet{QinEtal2022} comprehensively tested the conclusions of \citet{Matsubara2007}, validated that the Gaussian approximation can hold up to scales of $\abs{\vect{k}} \sim 1\ \ihmpc$ for one-point PDFs of both modulus and phases. Thereof the density field in Fourier space poses very interesting statistical features. Its Fourier mode (even in the non-linear regime) closely follows the Gaussian distribution, while phase correlation of two points is already non-negligible, and then bispectrum and trispectrum are apparently significant \citep[see also][]{MatarreseEtal1997, Scoccimarro2000, VerdeHeavens2001, GualdiEtal2021}. 

It appears that the Fourier transformation acts as a special type of Gaussianization, enabling us to understand the statistics of the comic density field in Fourier space at the one-point level.
This work goes further and aims to apply the idea to model the particular type of statistics about the ratio of two different cosmic fields $A$ and $B$ in Fourier space, namely $X_\kvec \equiv A_\kvec /B_\kvec$.

The mode-dependent growth function of the dark matter density field utilized in \citet{FalckEtal2021} falls into this category. In theory, the evolution of individual modes of the density field is intrinsically deterministic if actions other than gravitational force can be ignored \citep[e.g.][]{SahniColes1995, BernardeauEtal2002}. However, the complicated effects of mode coupling will make the growth have a distribution over different modes. Using multiple realizations of their {\tt Indra} simulation suits, \citet{FalckEtal2021} characterize the growth of Fourier modes of the density field by defining the mode-dependent growth function as $D_\kvec(z)=\abs{\delta_\kvec(z)/\delta_\kvec(z_i)}$. The Fourier-transformed density contrast $\delta_\kvec$ at a given redshift $z$ and the initial redshift $z_i$ of their simulation are used. They found that $D_\kvec(z)$ is stochastic over different realizations, and its distribution becomes wider at later epochs and higher $k=\abs{\kvec}$. The medians of $D_\kvec$ and the averages of $\ln D_\kvec$ were also shown on various scales and at different redshifts, but no analytical formulae are provided for the distribution and related statistics. Here, we will demonstrate that the analytical approximation to the one-point PDF of $D_\kvec$ can be derived based on the results of \citet{Matsubara2007}.

In principle, the Fourier mode of any type of nonlinear cosmic field, including momentum divergence, shall also obey the theorem of \citet{Matsubara2007}, as long as prerequisite conditions are satisfied. \citet{JenningsJennings2015} proposed a concept of stochastic growth rate to help develop perturbative theories of the redshift distortion effect, based on their numerical findings on the ratio of the velocity divergence field to the density field in Fourier space. This is another example of the ratio of cosmic fields in Fourier space, whose distribution could be examined and formulated. It is still challenging to conduct robust and accurate statistics of the volume-averaged velocity field from discrete samples deep into the nonlinear regime. Therefore, we instead consider using the momentum divergence field, since the momentum divergence could be easily measured from simulation \citep{Pan2020}. 

As will be shown later, for dark matter, the quotient of momentum divergence to density is a direct measure of the growth rate of the density field at any given moment $z$ and scale $\abs{\vect{k}}$. Thus, the mode growth rate defined in this way could be linked to the mode-dependent growth function $D_\kvec(z)$. But unlike the stochastic growth rate in \citet{JenningsJennings2015}, it could not be directly used for the redshift distortion effect. 

The entire paper is structured as follows. The definition and theoretical foundation are laid mainly in Section~\ref{sec:math}, which is then compared with the simulation data in Section~\ref{sec:sim}. Section~\ref{sec:zmean_model} is about the performance test of empirical models for the mean mode growth rate. This study ends with a summary and discussion in the last section.

\section{Theories on statistics of quotient of two complex Gaussian random variables}
\label{sec:math}
Starting from a more general expression of the distribution of two complex Gaussians, we summarize the major mathematical background. The results are then extended with two particular statistics, namely, the mode-dependent growth function and the mode growth rate. Their statistical meanings are stated and will be investigated in the following part of this work.
\subsection{Distribution of the ratio of two complex Gaussians}
\label{sec:general_math}
Without loss of generality, $A_\kvec$ and $B_\kvec$ are supposed counterparts in the Fourier space of two random fields $A$ and $B$ of any kind, then $X_\kvec$ is defined as the quotient of the two complex random variables, $X_\kvec=A_\kvec/B_\kvec$. Calculating the distribution of $X_\kvec$ requires complete knowledge of the joint distribution of $A_\kvec$ and $B_\kvec$. Practically, the closed-form analytical expression is intractable, except for a few cases. Fortunately, if $A_\kvec$ and $B_\kvec$ are both distributed as Gaussian, analytical formulae exist in the literature, which we reproduce here as the basis of our work.

Let ${\bf q}=[A_\kvec, B_\kvec]^T$ be a $2\times 1$ complex Gaussian random vector, there are the mean of the vector ${\bf u}=\avg{{\bf q}} =[u_A, u_B]^T$ and the covariance matrix 
\begin{eqnarray}
\avg{ ({\bf q}-{\bf u})({\bf q}-{\bf u})^H}  = \left[
\begin{array}{cc}
\sigma_A^2 & r \sigma_A \sigma_B \\
r^*  \sigma_A \sigma_B & \sigma_B^2 
\end{array} \right]\ ,
\end{eqnarray}
$\avg{\ldots}$ means the average, $(\ldots)^*$ is the conjugate, $(\ldots)^T$ and $(\ldots)^H$ refer to the transpose and conjugate transpose respectively, and r is the parameter of the correlation coefficient. Let $X_\kvec= A_\kvec/B_\kvec=X_R+i X_I$, \citet{LiHe2019} derive that the joint PDF of $X_R$ and $X_I$ can be written as
\begin{equation}
p(X_R, X_I)= \frac{e^{-c_2}}{ c_1 \pi \sigma_B^2 \gamma^2(X_\kvec)} e^{\frac{\lambda^2(X_\kvec)}{\gamma(X_\kvec)} }\left[ 1+\frac{\lambda^2(X_\kvec)}{\gamma(X_\kvec)} \right]\ ,
\label{eq:ratio_general}
\end{equation}
in which
\begin{align*}
c_1 &= \sigma_A^2 (1-\abs{r}^2), \notag \\
c_2 &= \abs{a_2}^2/c_1 + \abs{u_B}^2/\sigma^2_B, \notag \\
\gamma(X_\kvec)  &= \abs{ X_\kvec - a_1}^2 / c_1+1/\sigma_B^2, \notag \\
\lambda(X_\kvec) &= \abs{a_2(X_\kvec- a_1)^*/c_1+u_B/\sigma^2_B}, \notag
\end{align*}
with $a_1=r \sigma_A/\sigma_B$ and $a_2=u_A-a_1u_B$. Note that the formula is general and is not restricted to the case of independently distributed $A_\kvec$ and $B_\kvec$.

The mean of $X_\kvec$, if $u_B\neq 0$, is given by
\begin{equation}
\avg{X_\kvec} =
\frac{u_A}{u_B}+ \left( r\frac{\sigma_A}{\sigma_B}-\frac{u_A}{u_B} \right)  e^{-\abs{u_B}^2/\sigma_B^2} \ ,
\label{eq:mean_general}
\end{equation}
and if $u_B=0$
\begin{equation}
\avg{X_\kvec}= r \frac{\sigma_A}{\sigma_B} \ ,
\label{eq:mean_ub0}
\end{equation}
while higher-order moments $\avg{\abs{X_\kvec}^{n\geq 2}}$ do not exist unless $\abs{r}=1$ \citep{Wu2019, LiHe2019}.

If $A_\kvec$ and $B_\kvec$ are independent, then $r=0$ and the joint PDF reduces to 
\begin{equation}
p(X_R, X_I) = \frac{ \sigma_A^2 \sigma_B^2 e^{ -\left( \frac{\abs{u_A}^2}{\sigma_A^2} + \frac{\abs{u_B}^2}{\sigma_B^2}  \right)} }{\pi ( \abs{X_\kvec}^2 \sigma_B^2+\sigma_A^2)^2} e^{\kappa(X_\kvec)} [1+\kappa(X_\kvec)]
\end{equation}
with 
\begin{equation}
\kappa(X_\kvec)= \frac{\abs{ X_\kvec^* u_A/\sigma_A^2 + u_B/\sigma_B^2}^2 }{(\abs{X_\kvec}^2 /\sigma_A^2+1/\sigma_B^2)} \notag
\end{equation}
\citep[e.g.][]{Pham-GiaEtal2006, NadimiEtal2018}.

A more special case is $u_A=u_B=0$, the joint PDF is in the form of
\begin{equation}
p(X_R, X_I)=\frac{1-\abs{r}^2}{\pi \left[
\abs{X_\kvec}^2\frac{\sigma_B}{\sigma_A}+\frac{\sigma_A}{\sigma_B}-2 (r_R X_R + r_I x_I)
\right]^{2}}\ ,
\label{eq:ratio_used}
\end{equation}
in which $r=r_R+i r_I$ \citep{BaxleyEtal2010}. At this stage, we have derived the main equation, which will be extended to two particular statistics in the context of structure formation in the following part of this study.

\subsection{The mode-dependent growth function}
\label{sec:gf_math}
For generality, we slightly extended the notion of \citet{FalckEtal2021}, by defining a complex mode-dependent growth function ${\mathcal D_\kvec}$,
\begin{equation}
X_\kvec = \frac{\delta_\kvec(z_1)}{\delta_\kvec(z_2)} \equiv \mathcal{D}_\kvec(z_1,z_2)\ , 
\end{equation}
where $\delta_\kvec(z)$ is the Fourier counterpart of the dimensionless density contrast $\delta=\rho/\avg{\rho}-1$ at redshift $z$, with $\rho$ denoting the density and $z_2 > z_1$. If we denote $\mathcal{D}_\kvec(z_1,z_2) = D_\kvec(z_1,z_2) e^{i\phi_D}$, the amplitude $D_\kvec=\abs{\mathcal D_\kvec}$ is commonly used and is the target function that we want to analyze. 
As formerly mentioned in Section~\ref{sec:intro}, the nonlinear field of $\delta_\kvec(z)$ in the Fourier space could be approximately treated as Gaussian distributed. This generally guarantees the usage of Equation~(\ref{eq:ratio_general}) to model $p(D_\kvec, \phi_D)$.

Let the density power spectrum at $z$ be $P_z(\kvec, z)\equiv \avg{\delta_\kvec(z)\delta^*_\kvec(z)}$, and let the density cross-spectrum at two different redshifts be $P_{z_1,z_2}(\kvec, z_1, z_2)\equiv \avg{\delta_\kvec(z_1)\delta^*_\kvec(z_2)}$. 
With the consideration that the density field is statistically isotropic and homogeneous, $\delta_{\kvec}$ satisfies $\avg{\delta_\kvec}=0$ and $P_{z_1,z_2}=P^*_{z_1,z_2}$. This exactly corresponds to the case of Equation~(\ref{eq:ratio_used}), but with the correlation coefficient $r_R=P_{z_1,z_2}/\sqrt{P_{z_1}P_{z_2}}$ and $r_I=0$. 
If we further let $t=\sqrt{P_{z_1}/P_{z_2}}$, Equation~(\ref{eq:ratio_used}) proceeds to another form
\begin{eqnarray}
\label{eq:Dk_pdf}
&p(D_\kvec, \phi_D) &=  \frac{1-r^2}{\pi}\frac{D_\kvec t^2}{ \left( 
D^2_\kvec  + t^2 - 2 D_\kvec  r t \cos\phi_D \right)^2 }\ ,  \nonumber \\
&p(D_\kvec) &= \frac{2 D_\kvec t^2 (D_\kvec^2+t^2) (1-r^2)}{ \left[ (D_\kvec^2+t^2)^2 - 4 D_\kvec^2 t^2  r^2 \right]^{3/2}}\ ,  \nonumber \\
&p(\phi_D) & = \frac{1-r^2}{2\pi (1- r^2 \cos^2\phi_D)^{3/2} } 
\{  \sqrt{1-r^2\cos^2\phi_D}      \nonumber \\ 
 & & +  [\pi -  \arccos (r\cos\phi_D)] r \cos \phi_D  \} \ ,
\end{eqnarray}
with $r<1$ and $-\pi \leq \phi_D < \pi$.

With the results of Section~\ref{sec:general_math}, we could derive the mean $\avg{\mathcal D_\kvec}=rt $ and the median $D_{\kvec,1/2}=t$. What is important is that we have identified the meanings of two often used quantities. The first is that $P_{z_1,z_2}/P_{z_2}$ is the mean of the complex mode-dependent growth function ${\mathcal D_\kvec}$; the second is that $\sqrt{P_{z_1}/P_{z_2}}$ is the median value of $D_{\kvec}$.

Unfortunately, the mean $\avg{D_\kvec}$ does not have a simple analytic form and could only be calculated with numerical integration. In real applications, one could always numerically compute an averaged $D_\kvec$ over many simulation realizations. However, the mean value of the logarithm of $D_\kvec$ used in \citet{FalckEtal2021}, $\avg{\ln D_\kvec}$, exists. Under the Gaussian approximation, $\ln\abs{\delta_\kvec}$ obeys the dubbed log-Rayleigh distribution, in which the expectation is
\begin{equation}
\avg{\ln \abs{\deltak}}= \ln \sqrt{\frac{P_\delta}{2}} + \frac{\ln 2}{2} -\frac{C}{2}\ ,
\label{eq:lg_mean}
\end{equation}
where $C$ is the Euler constant defined by $C=- \int_0^{+\infty} e^{-x}\ln x \rmd x$ \citep{RivetEtal2007}. Subsequently, $\avg{\ln D_\kvec}$ could be related to the density power spectrum as 
\begin{equation}
\avg{\ln D_\kvec} = \avg{\ln \abs{\deltak(z_1)}} -\avg{\ln \abs{\deltak(z_2)}} = \frac{1}{2}\ln \frac{P_{z_1}}{P_{z_2}}\ ,
\end{equation}
which can be used straightforwardly to interpret the results of \citet{FalckEtal2021}.

\subsection{A minor note about possible applications to bias}
If we go beyond the dark matter density field by considering the density field of biased tracers in the Fourier space $\delta_g(\kvec)$, which has a complex bias function as $\widetilde{b}_\kvec=\delta_g(\kvec)/\delta_m(\kvec)$ relative to dark matter $\delta_m(\kvec)$, mathematical results in Section~\ref{sec:gf_math} can be essentially adopted after the replacements of $\deltak(z_1)\rightarrow \delta_g(\kvec)$ and $\deltak(z_2)\rightarrow \delta_m(\kvec)$. Therefore, the ordinary bias function $b_o=\sqrt{P_g/P_m}=\sqrt{\avg{\delta_g\delta_g^*}/\avg{\delta_m \delta_m^*}}$ is the median value of $\abs{\widetilde{b}_\kvec}$. While another commonly cited bias function $b_x=P_{gm}/P_m=\avg{\delta_g\delta_m^*}/\avg{\delta_m \delta_m^*}$ is the average of $\widetilde{b}_\kvec$, with denoting $P_{gm} = \avg{\delta_g\delta_m^*}$ as the cross-spectrum of the two field. The difference between $b_o$ and $b_x$ vanishes in the limit of $P_{gm}/\sqrt{P_gP_m}\rightarrow 1$, that is the case when bias shows little stochasticity. 

The number density of biased traces such as halos and galaxies is usually much lower than that of dark matter in simulations. Understanding and devising methods to suppress discreteness effects on $\delta_g(\kvec)$ is not an easy task. Therefore, we will leave further exploration of biased density fields for future research and concentrate on the distribution of dark matter only. 

\subsection{The mode growth rate}
\label{sec:gr_math}

In analogy with \citet{JenningsJennings2015}, we start from the continuity equation, which is valid for dark matter as long as dark matter does not annihilate significantly,
\begin{equation}
 a \frac{\partial \delta({\mathbf r}, \tau)}{\partial \tau}+\nabla \cdot\{[1+\delta({\mathbf r}, \tau)] {\mathbf v}({\mathbf r}, \tau)\}=0 \ ,
\end{equation}
in which $a=1/(1+z)$ with $z$ being the redshift corresponding to cosmic time $\tau$, $\delta$ is the dark matter density contrast at position ${\mathbf r}$, and ${\bf v}$ labels the peculiar velocity. In Fourier space, it turns out to be
\begin{equation}
   \frac{\partial \deltak(a)}{\partial \ln a}=f \psi_\kvec(a) \ , 
\label{eq:cont_complex}
\end{equation}
in which $\psi_\kvec$ (shorthand for $\psi(\kvec)$) is the Fourier transform of the momentum divergence $\psi(\vect{r}) \equiv - \nabla \cdot \left[(1+\delta) {\bf v}(\vect{r}) \right]/ Haf$, where $H$ is the Hubble parameter, and $f=\rmd \ln D/\rmd \ln a$ is the linear growth rate defined with respect to the linear growth function $D$. Unlike $D_\kvec$, the linear growth function is scale independent and is defined as the amplitude factor of the density contrast $\delta$ relative to that at the present time.

On the other hand, if we denote $\deltak=\abs{\deltak} e^{i\phik}$, then
\begin{equation}
\frac{1}{\deltak}    \frac{\partial \deltak}{\partial \ln{a}} 
= \frac{1}{\abs{\deltak}} \frac{\partial \abs{\deltak}}{\partial \ln{a}} + 
    i \frac{\partial \phik}{\partial \ln{a}} \ ,
\label{eq:derivative}
\end{equation} 
of which the real part is only about the growth of the moduli and the imaginary component is solely about phases. 
Let $X_\kvec \equiv \psi_\kvec/\deltak$, combining Equation~(\ref{eq:cont_complex}) with Equation~(\ref{eq:derivative}) yields the following, 
\begin{equation}
X_\kvec=\frac{1}{\deltak} \frac{\partial \deltak}{\partial \ln D}=\Delta'+i \phi'\ .
\end{equation}
$\Delta'$ and $\phi'$ refer to the real and imaginary parts of $X_\kvec$, respectively, 
\begin{equation}
 \Delta' \equiv X_R = \frac{ \partial\ln \abs{\deltak}}{\partial \ln D}\ , \quad 
 \phi' \equiv X_I = \frac{\partial \phik}{\partial \ln D} \ .
\label{eq:z_def}
\end{equation}
Here, $a$ is replaced by the linear growth function $D$ as the time variable. 

It is apparent that $X_\kvec$ effectively becomes the {\em mode growth rate} of dark matter's density field, of which $\Delta'$ is the {\em mode's amplitude growth rate} and $\phi'$ stands for the {\em mode's phase growth rate}. 
In the linear regime $\psi_\kvec=\deltak$, $\Delta'=1$ and $\phi'=0$, the amplitudes will grow linearly with $D$ while the phases will remain invariant. However, when gravitational nonlinearity increases,  $\psi_\kvec \neq \deltak$, it is expected that stochasticity in $\Delta'$ and $\phi'$ will become stronger due to the effects of mode coupling and multi-streaming. 

Exact modeling of the distribution of $X_\kvec$ requires complete knowledge of the joint distribution of $\deltak$ and $\psi_\kvec$. Again, based on the results of \citet{Matsubara2007}, if $\psi_\kvec$ closely follows the Gaussian distribution in a similar way as $\deltak$ does, then the results of Section~\ref{sec:general_math} can be used to model $p(\Delta',\phi')$. 

Since $\avg{\deltak}=\avg{\psi_\kvec}=0$, after denoting
\begin{align*}
& P_{\delta}(\vect{k}) = \avg{\deltak \deltak^*}, \, P_{\psi}(\kvec) = \avg{\psi_\kvec \psi^*_\kvec}, \,
 P_{\delta\psi} = \avg{\deltak\psi_\kvec^*}=\avg{\deltak^*\psi_\kvec} \\
\end{align*}
and 
\begin{align*}
& \alpha \equiv P_{\delta\psi}/\sqrt{P_{\delta} P_{\psi}}, \quad \beta \equiv \sqrt{P_\psi/P_\delta}, 
\end{align*}
the one-point PDFs of $X_\kvec$, $\Delta'$ and $\phi'$ can be derived from Equation~(\ref{eq:ratio_used}) without difficulty,
\begin{eqnarray}
&p(\Delta',\phi') & =\frac{1}{\pi}\frac{\beta^2(1-\alpha^2)}
{\left[ (\Delta' - \avg{\Delta'})^2 + \phi'^2 + \beta^2(1-\alpha^2)\right]^2} \nonumber \\
&p(\Delta')&= \frac{1}{2} \frac{\beta^2 (1-\alpha^2)}
{ \left[ (\Delta'-\avg{\Delta'})^2+ \beta^2(1-\alpha^2) \right]^{3/2}} \nonumber \\
&p(\phi')&= \frac{1}{2} \frac{\beta^2 (1-\alpha^2)}
{\left[ \phi'^2 +\beta^2(1-\alpha^2) \right]^{3/2}}\ ,
\label{eq:z_pdf}
\end{eqnarray}
both $p(\Delta')$ and $p(\phi')$ are actually the Student's t-distribution with 2 degrees of freedom.

A nice property of defining the mode growth rate with Equation~(\ref{eq:z_def}) is that the first-order moment $\avg{X_\kvec}$ is well determined, 
\begin{eqnarray}
&\avg{\Delta'} &= \avg{X_R}=\left\langle\frac{\psi_\kvec}{\deltak} \right\rangle=
\alpha \beta=\frac{P_{\delta\psi}}{P_{\delta}}\ , \nonumber \\
&\avg{\phi'}& = \avg{X_I}=0 \ .
\label{eq:z_mean}
\end{eqnarray}
In fact, Equation~(\ref{eq:z_mean}) also points out that $\avg{X_\kvec}$ can be estimated by the ratio of two nonzero power spectra, free of the possible numerical catastrophe when $\abs{\deltak}\sim 0$.  

\subsection{Links among mode-dependent growth function, mode growth rate and power spectrum}
\label{sec:zmean_pk}

The amplitude of the mode-dependent growth function at redshift $z_1$ with respect to $z_2$ ($z_2 > z_1$) could actually be related to $\Delta'$ through
\begin{equation}
\ln D_\kvec(z_1, z_2) = \ln \frac{\abs{\deltak(z_1)}}{\abs{\deltak(z_2)}} =\int_{D(z_2)}^{D(z_1)} \Delta' \rmd \ln D\ .
\end{equation}
Taking the average in the Gaussian ansatz then leads to
\begin{equation}
\frac{P_\delta(z_1)}{P_\delta(z_2)}=e^{2\avg{\ln D_\kvec(z_1, z_2)}}= \exp\left(2 \int_{D(z_2)}^{D(z_1)} \avg{\Delta'} \rmd \ln D\right) \ ,
\label{eq:pkevolution}
\end{equation}
or
\begin{equation}
\rmd \ln P_\delta /\rmd \ln D = 2 \avg{\Delta'}\ .
\label{eq:zmean_pk}
\end{equation} 
This means that $\avg{\Delta'}$ is a measure of the logarithmic growth rate of the nonlinear density power spectrum, as long as the Gaussian approximation to $\deltak$ and $\psi_\kvec$ holds its effectiveness. 

\section{Comparison with N-body simulation}
\label{sec:sim}
In Section~\ref{sec:math}, the analytical expressions of the distributions of the mode-dependent growth function and the mode growth rate are derived under the assumption that the density field $\deltak$ and the momentum divergence field $\psi_\kvec$ follow the Gaussian distribution. In this section, simulation data are used to examine whether the prerequisite conditions are met and then the analytical formulas of two statistics are compared.
\subsection{The simulation and numerical issues}
\label{sec:sim_data}
In this work, the {\tt Pangu} simulation in \citet{LiEtal2012} is recruited. The dark matter-only simulation is conducted with a memory-optimized version of {\tt GADGET2} \citep{Springel2005}. The cosmological parameters are $\Omega_m=0.26$, $\Omega_b=0.044$, $\Omega_\Lambda=0.74$, $h=0.71$ and $\sigma_8=0.8$. The simulation uses $N = 3072^3$ dark matter particles in a cubic periodic box with $L = 1\ h^{-1}{\rm Gpc}$ on one side. Each particle has a mass of $2.48915\times 10^9\ \msun$ and the Plummer-equivalent force softening length is $7\ \lhkpc$. The simulation starts from an initial redshift of $z=127$. Since we are mainly interested in regimes with considerably developed non-Gaussianity, about 13 snapshots at redshifts from $z=2$ to 0 are selected. The mass resolution of the simulation supports accurate measurement of power spectrum at scales $1<k\lesssim 10\ \ihmpc$, meanwhile its box size can provide sufficient numbers of modes at large scales $k<0.1\ \ihmpc$ and makes the systematics in modes of short wavelengths induced by the absence of long-wavelength modes negligible \citep[e.g.][]{CrocceScoccimarro2006, TakahashiEtal2008}.

The numerical fast Fourier transformation (FFT) is implemented with the {\tt FFTW3} package \citep{FFTW05}. To obtain accurate measurements of Fourier modes of density, momentum divergence, and related power spectra, the prescription of \citet{YangEtal2009} and \citet{Pan2020} is adopted to compensate the effects of aliasing and smoothing. The third-order orthogonalized Battle-Lemarié spline function is used when assigning dark matter particles upon FFT grids. But aliasing cannot be completely removed, and numerical experiments indicate that if $1\%$ precision is required, it is safe to use Fourier modes on scales $k< 0.67\ k_{N}$ (Appendix~\ref{app:a}), where $k_N$ is the Nyquist frequency. 

The quality of the numerical measurement is further controlled by applying criteria based on shot noise. We restrict our exploration to the scale range within which the influence of the power spectrum of the shot noise is less than $1\% $. Additionally, the number of FFT grids is set by the requirement that the mean number of particles in a grid cell should be greater than one. Therefore, the largest number of FFT grids used in this work is $2048^3$. By applying these conservative rules, the influence of discreteness becomes minuscule; henceforth, we do not make any shot noise correction unless specified otherwise. 

The last point we want to address here is that in this work only one simulation realization is used. Thus, the ergodicity principle is assumed when comparing simulations with theories. Specifically, statistics computed with all modes identical to $k=\abs{\kvec}$ from a single simulation are considered statistically equivalent to statistics of the mode at $\kvec$ over many realizations, with some reasonable fluctuations attributed to cosmic variance.

\subsection{One-point PDFs of $\deltak$ and $\psi_\kvec$}
\label{sec:sim_pdf}
\begin{table}
\centering
\caption{$k$ bin centres and their widths selected to calculate distribution functions.}
   \label{tab:kf_bins}
   \begin{tabular}{ccccccc}
	\hline
	$k$ & 0.05 & 0.10 & 0.30 & 0.60 & 1.00 & 2.00 \\  
	$\Delta k$ & 0.01 & 0.01 & 0.005 & 0.001 & 0.001 & 0.0005 \\
 	$N_{\mathrm{mode}}$ & 1431 & 5299 & 23246 & 18928 & 51089 & 99448 \\
        \hline
   \end{tabular}
   
   \begin{tablenotes}
     \footnotesize
      \item \textbf{Note.}  The number of modes $N_{\rm mode}$ are counted within $k\pm\Delta k$.
   \end{tablenotes}
\end{table}

\begin{figure*}[ht!]
\resizebox{\hsize}{!}{
\includegraphics{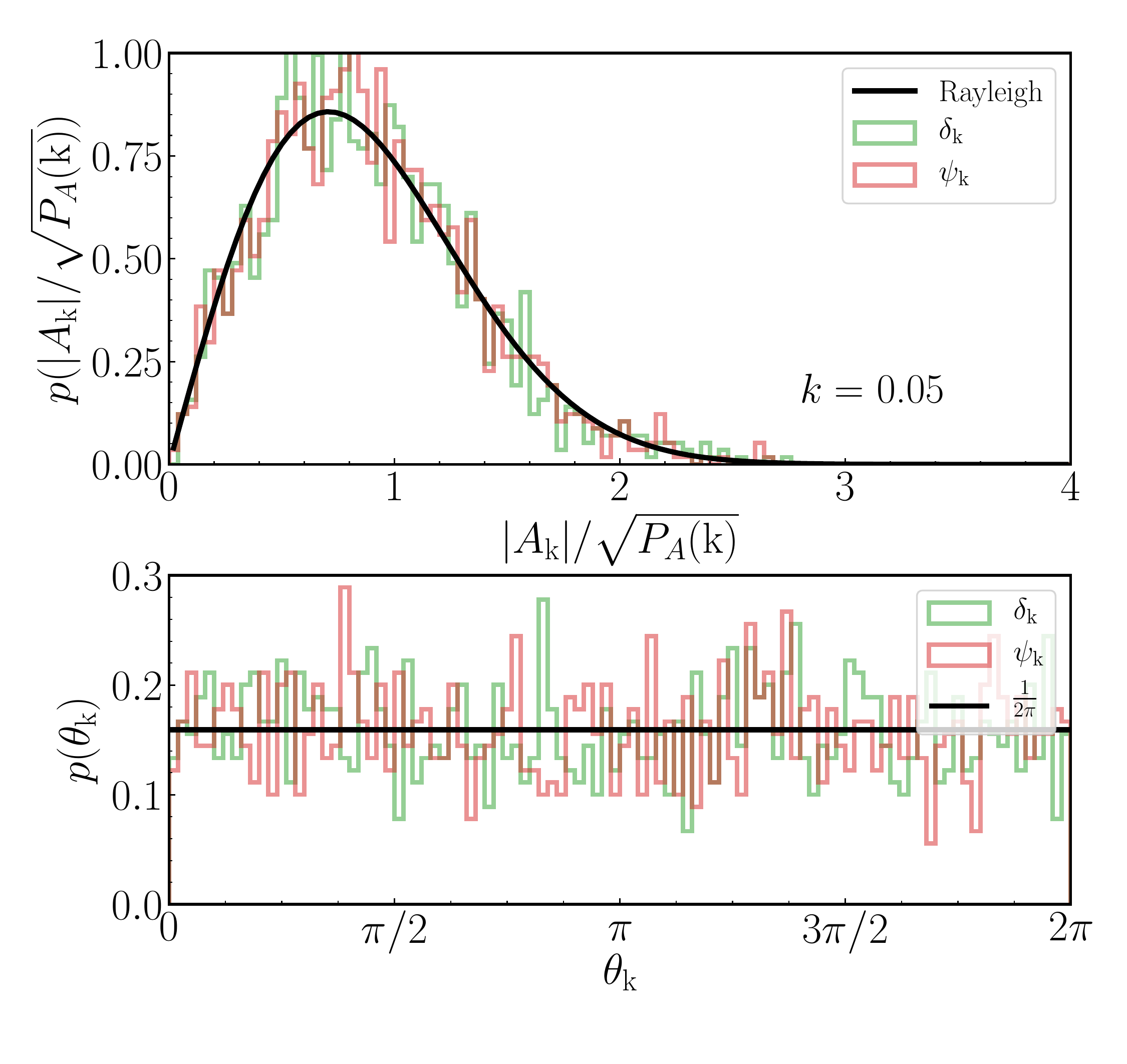} 
\includegraphics{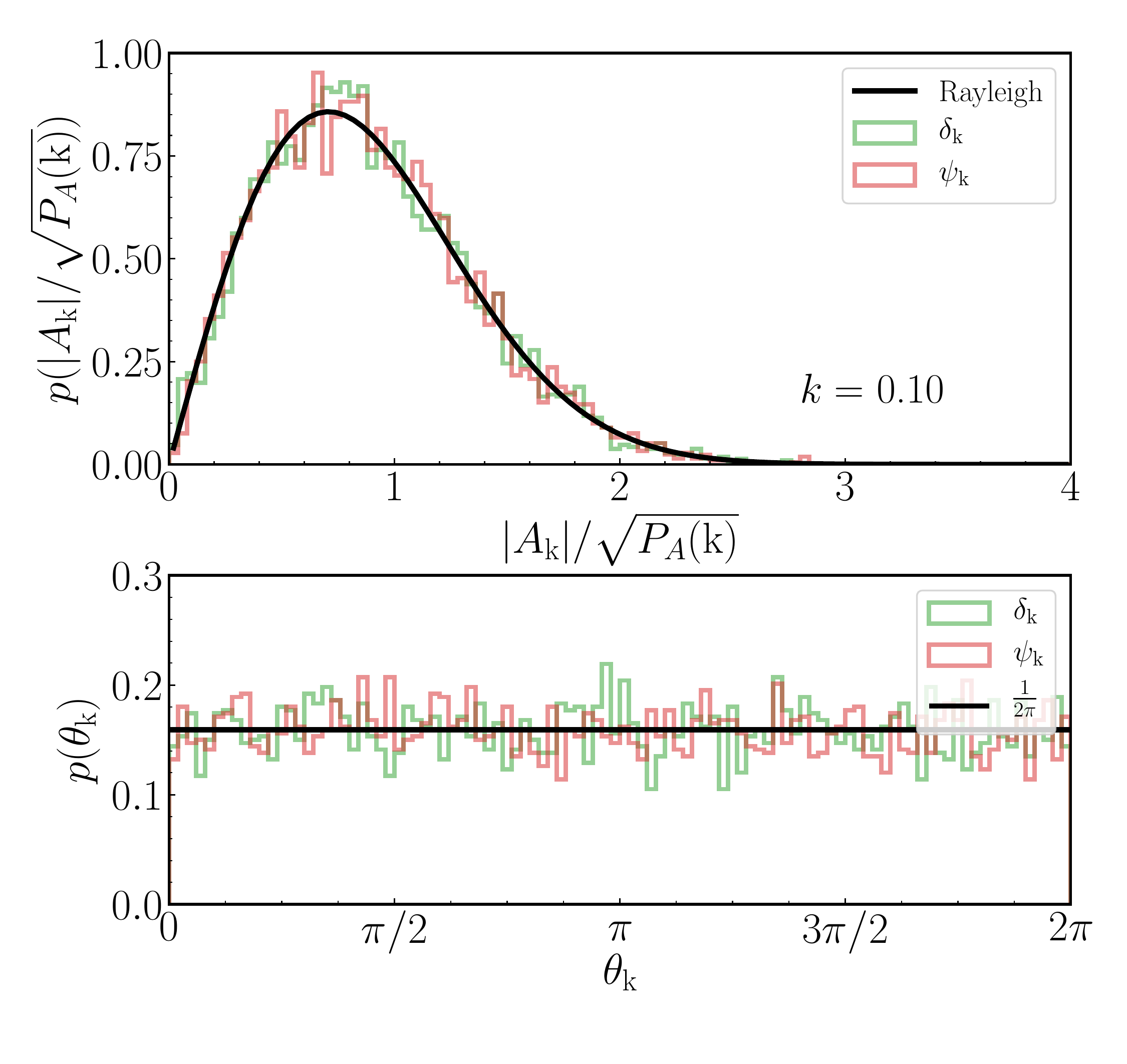}
\includegraphics{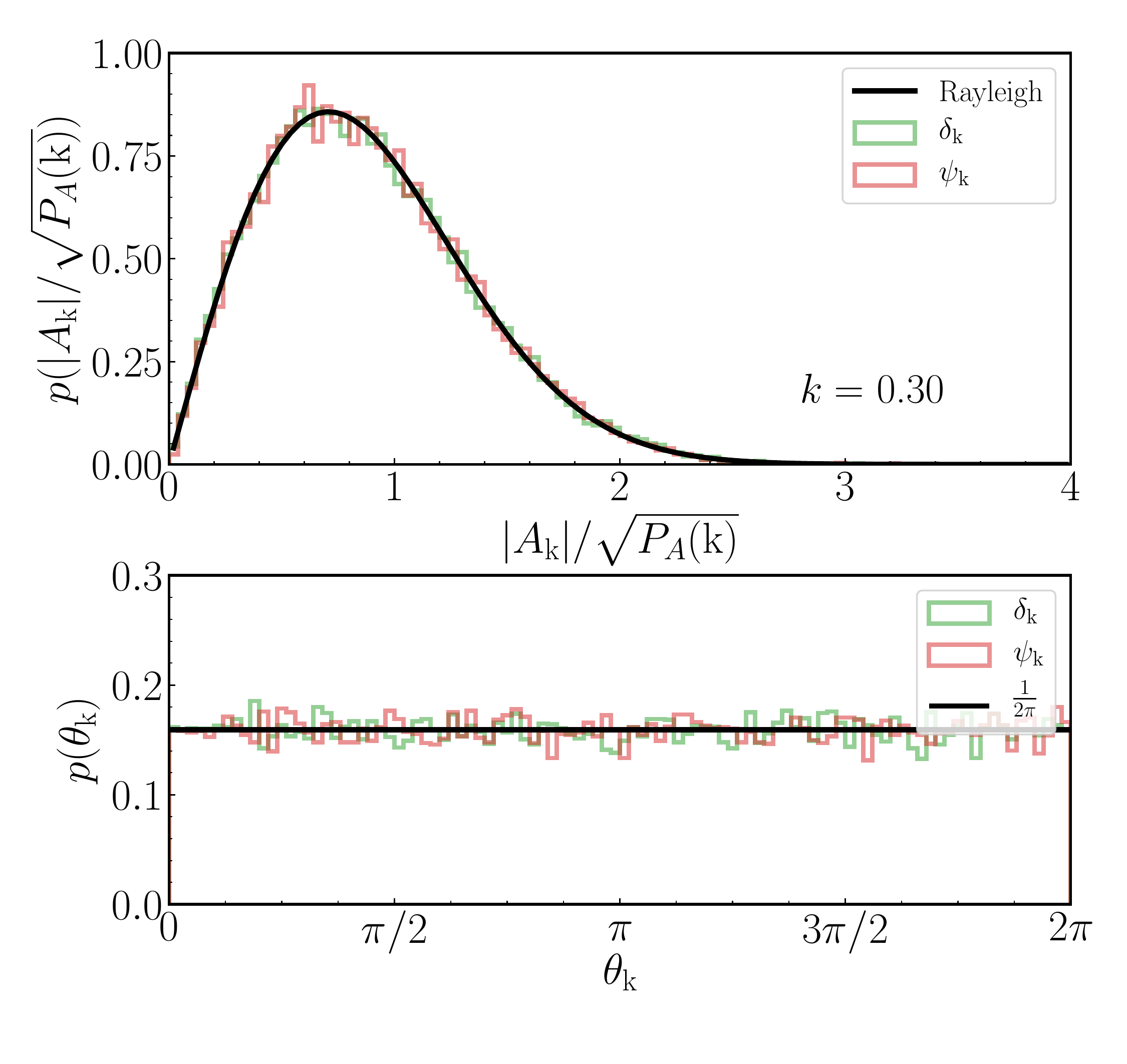} } \\
\resizebox{\hsize}{!}{
\includegraphics{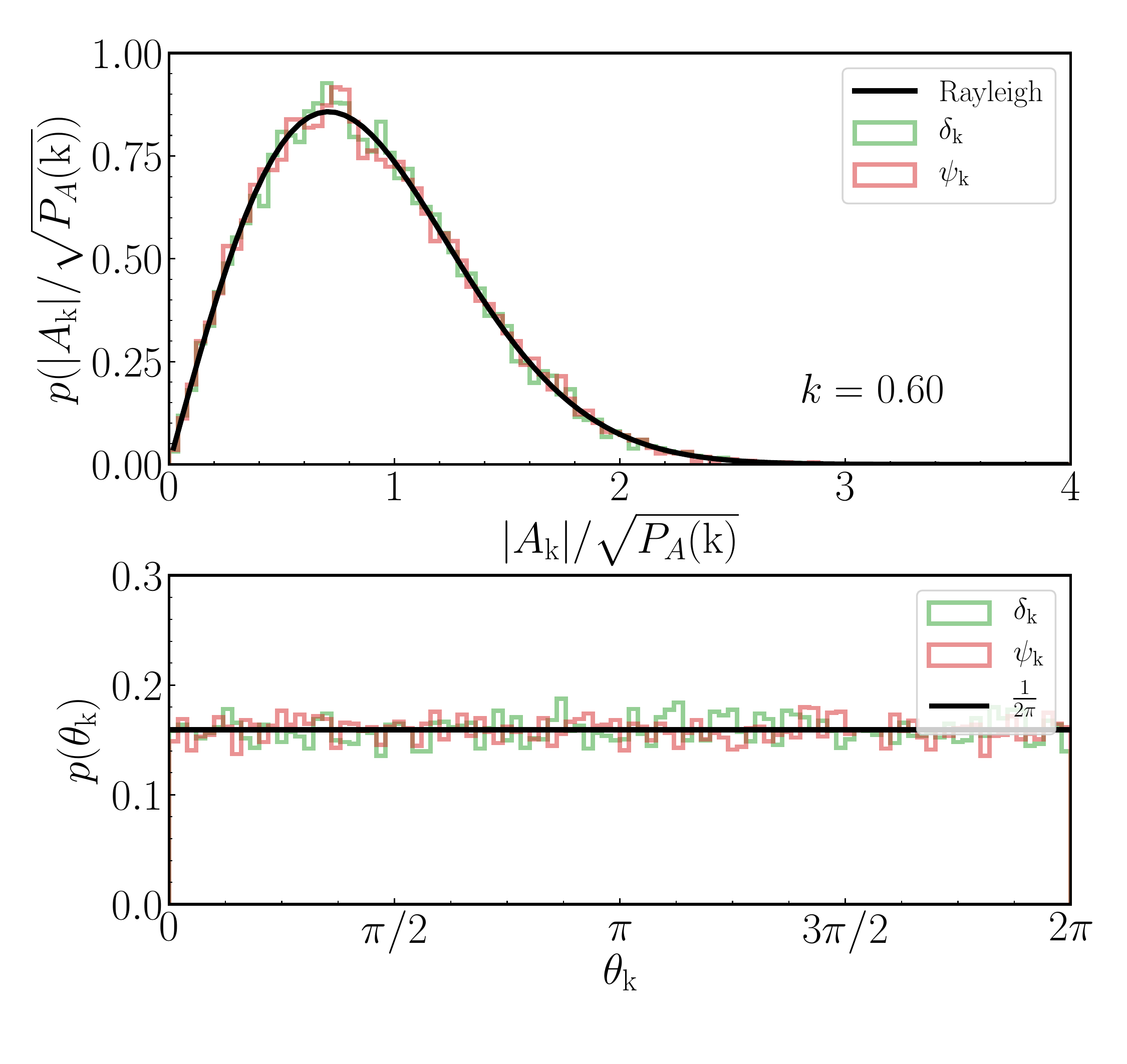}
\includegraphics{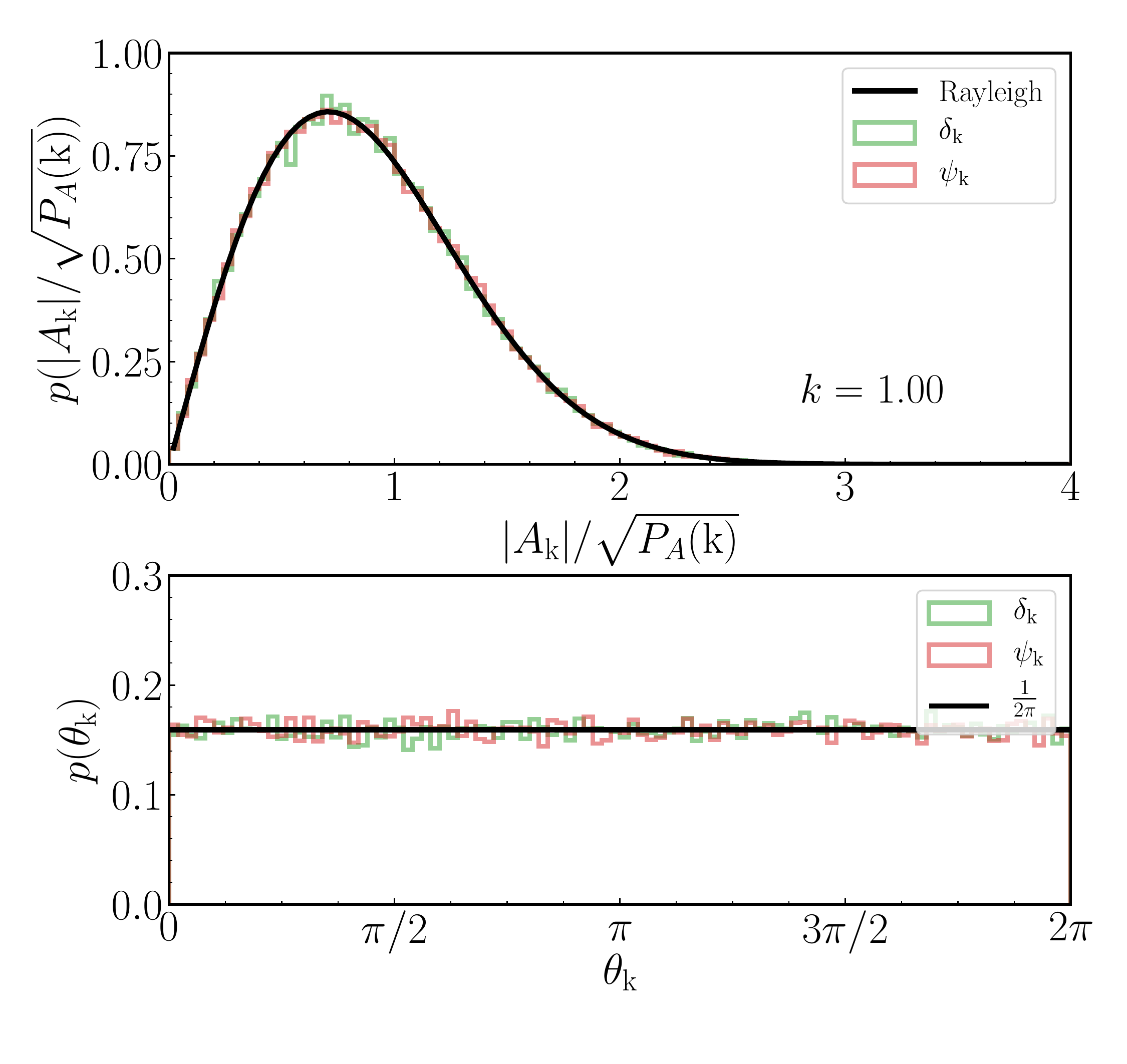} 
\includegraphics{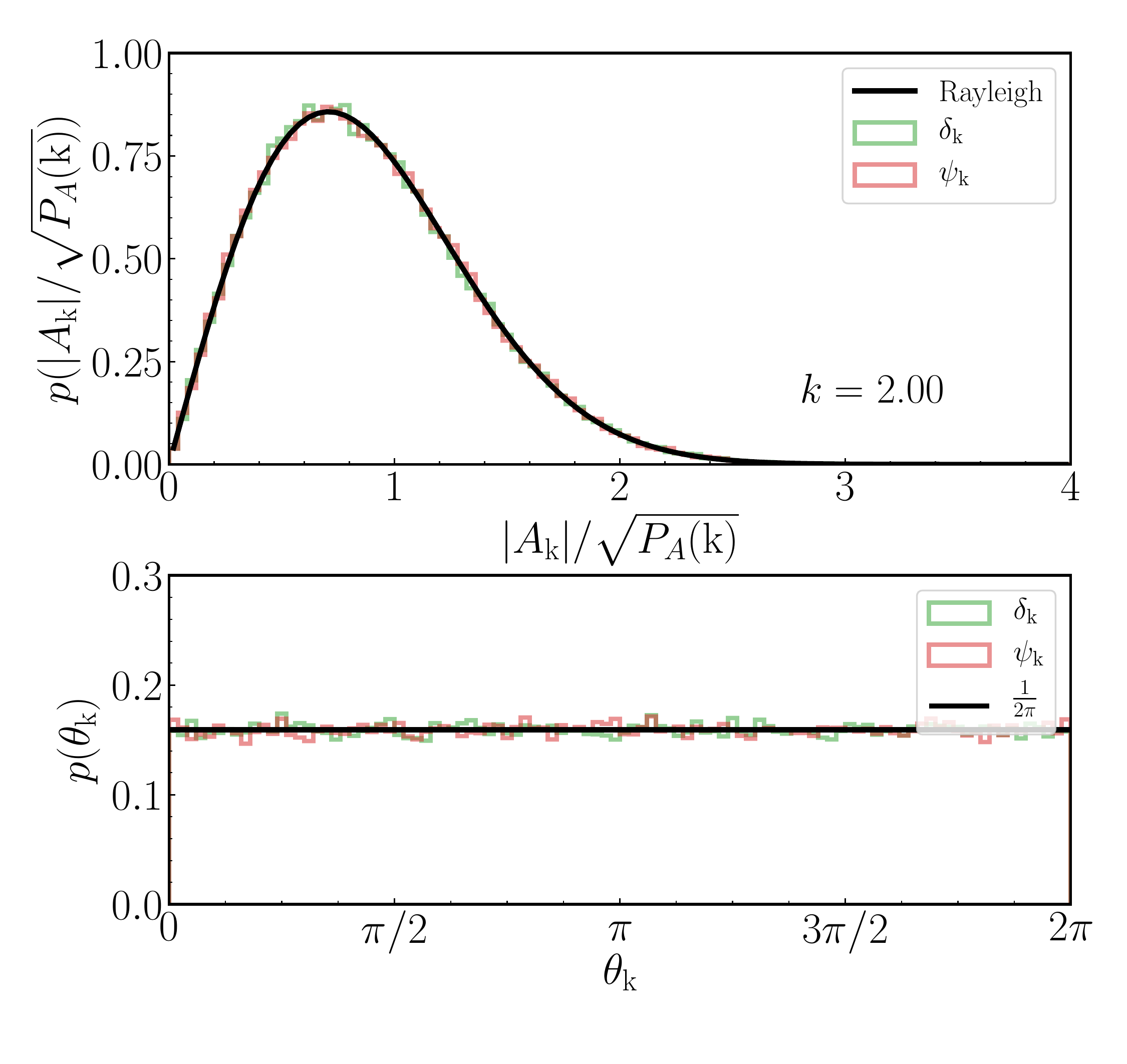}}
\caption{One-point PDFs of $\deltak$ and $\psi_\kvec$ of dark matter in the $z=0$ snapshot of the {\tt Pangu} simulation. The results are shown in six $k$ bins (Table~\ref{tab:kf_bins}). The symbol $A$ in the plot represents $\delta$ or $\psi$ accordingly.  Black solid lines are prediction by Equation~(\ref{eq:rayleigh}) with $P_\delta(\kvec)$ and $P_\psi(\kvec)$ estimated from the simulation.}
\label{fig:rayleigh}
\end{figure*}

If a complex random variable $A_\kvec=\abs{A_\kvec}e^{i\theta_\kvec}$ is Gaussian distributed with zero mean and variance $P_A=\avg{AA^*}$, its modulus $\abs{A_\kvec}$ will follow a Rayleigh distribution, and its phase $\theta_\kvec$ will be uniformly distributed,
\begin{equation}
 p(\abs{A_\kvec}, \theta_\kvec) \rmd \abs{A_\kvec} \rmd \theta_\kvec = 
\frac{\abs{A_\kvec} e^{-\abs{A_\kvec}^2/P_A(\kvec)} }{\pi P_A(\kvec)}  
\rmd \abs{A_\kvec}\rmd \theta_\kvec \ . 
 \label{eq:rayleigh}
\end{equation}
The $z=0$ snapshot of the {\tt Pangu} simulation, which has the strongest non-Gaussianity, is used to check up with the goodness of Gaussian approximation. The measured PDFs of $\deltak$ and $\psi_\kvec$ in together with predictions by Equation~(\ref{eq:rayleigh}) in six $k$ bins (Table~\ref{tab:kf_bins}) are shown in Figure~\ref{fig:rayleigh}, the agreement of the Gaussian model with the simulation is obvious. To quantify the deviations from Gaussianity, the skewness and kurtosis of the real and imaginary parts of $\deltak$ and $\psi_\kvec$ are calculated and shown in Figure~\ref{fig:dist_skew_kurt}. 

{As expected, across wide scale ranges from quasi-linear to strongly nonlinear regime, skewness and kurtosis are consistent with a Gaussian distribution. To demonstrate the contribution of finite mode numbers to the skewness and kurtosis, in each $k$ bin, we generate 1000 random Gaussian realizations. The sample size and the Gaussian parameters of each realization are kept the same as the simulation results. Then the skewness and kurtosis are extracted for each realization, and the mean and 1-$\sigma$ scatter of them are estimated. To reduce redundancy, only the results for the real parts of $\deltak$ and $\psi_\kvec$ are presented in Figure~\ref{fig:dist_skew_kurt}, shown as blue and range bands. The non-Gaussianity in one-point PDFs of $\deltak$ and $\psi_\kvec$ seems to be hardly associated with the strength of nonlinearity.

\begin{figure}
\plotone{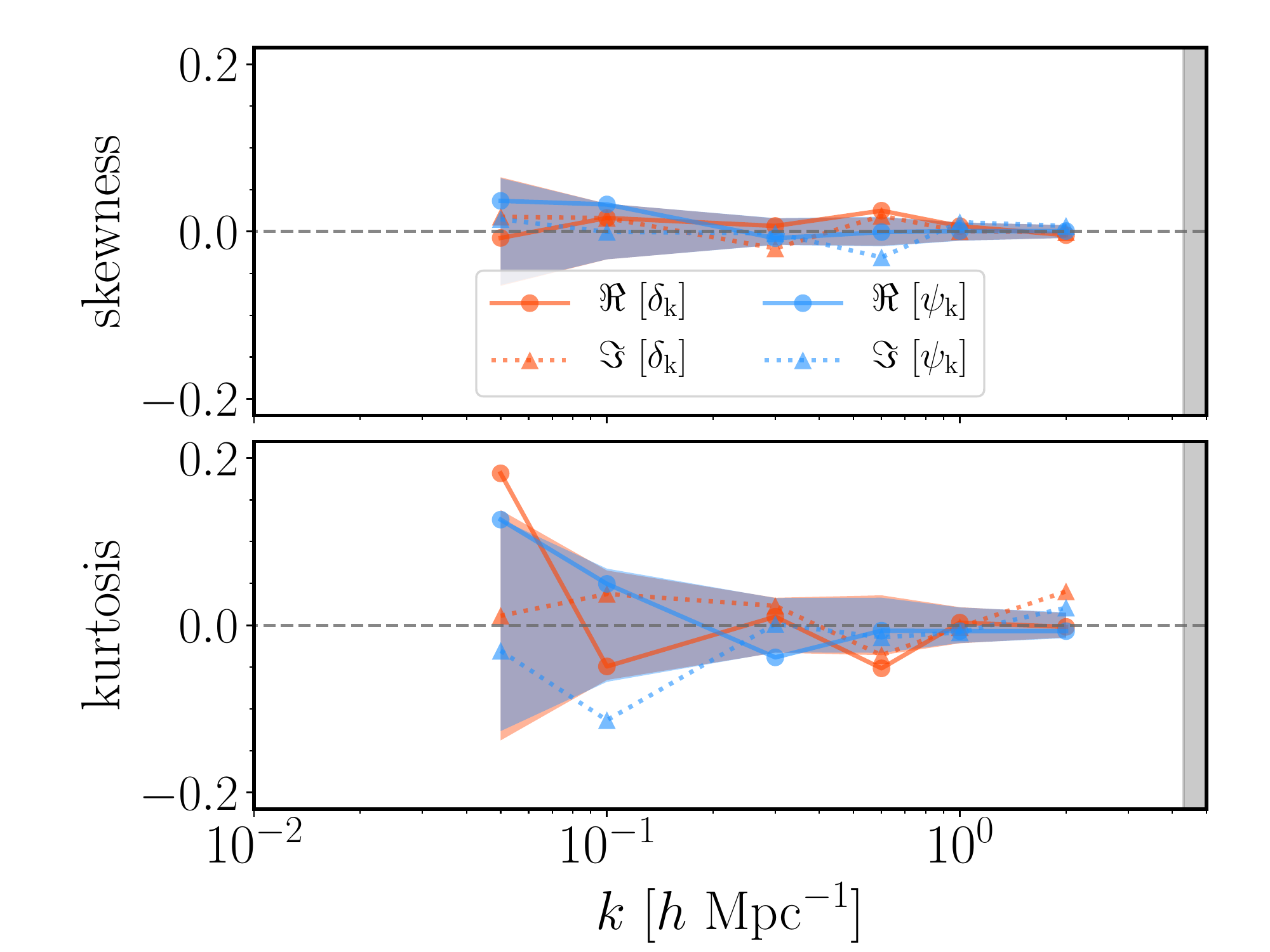}
\caption{Skewness and kurtosis of the real and imaginary parts of $\deltak$ and $\psi_\kvec$. Filled circles connected with solid lines are about the real parts, filled triangles linked with dashed lines are of the imaginary parts. For perfect Gaussian distribution, skewness and kurtosis are zeroes as denoted by those horizontal black dashed lines. The blue and orange bands are the 1$\sigma$ scatters estimated from 1000 Gaussian realisations in each $k$ bin. Only the real parts of $\deltak$ and $\psi_\kvec$ are shown for 
clearness. The shaded area at the right end is the regime where $k>0.67\ k_N$.}
\label{fig:dist_skew_kurt}
\end{figure}

Consequently, the conclusion is that the one-point distributions of $\deltak$ and $\psi_\kvec$ can be modeled by Equation~(\ref{eq:rayleigh}) remarkably well, even though the density field is highly non-Gaussian due to gravitational instability \citep[][]{FeldmanEtal2001,SefusattiEtal2006}. The results of the density field are also consistent with those of \citet{FalckEtal2021} and \citet{QinEtal2022}. What is new and of great importance is that, for the first time, the momentum divergence field of dark matter is verified to be highly Gaussianized at one-point level after the Fourier transformation. This extends the effectiveness of the results of \citet{Matsubara2007} and practically fulfills a crucial prerequisite condition for applying the theories in Section~\ref{sec:gr_math} to model the mode growth rate. 

\subsection{Distribution of mode-dependent growth function}

\begin{figure*}
\resizebox{\hsize}{!}{
\includegraphics{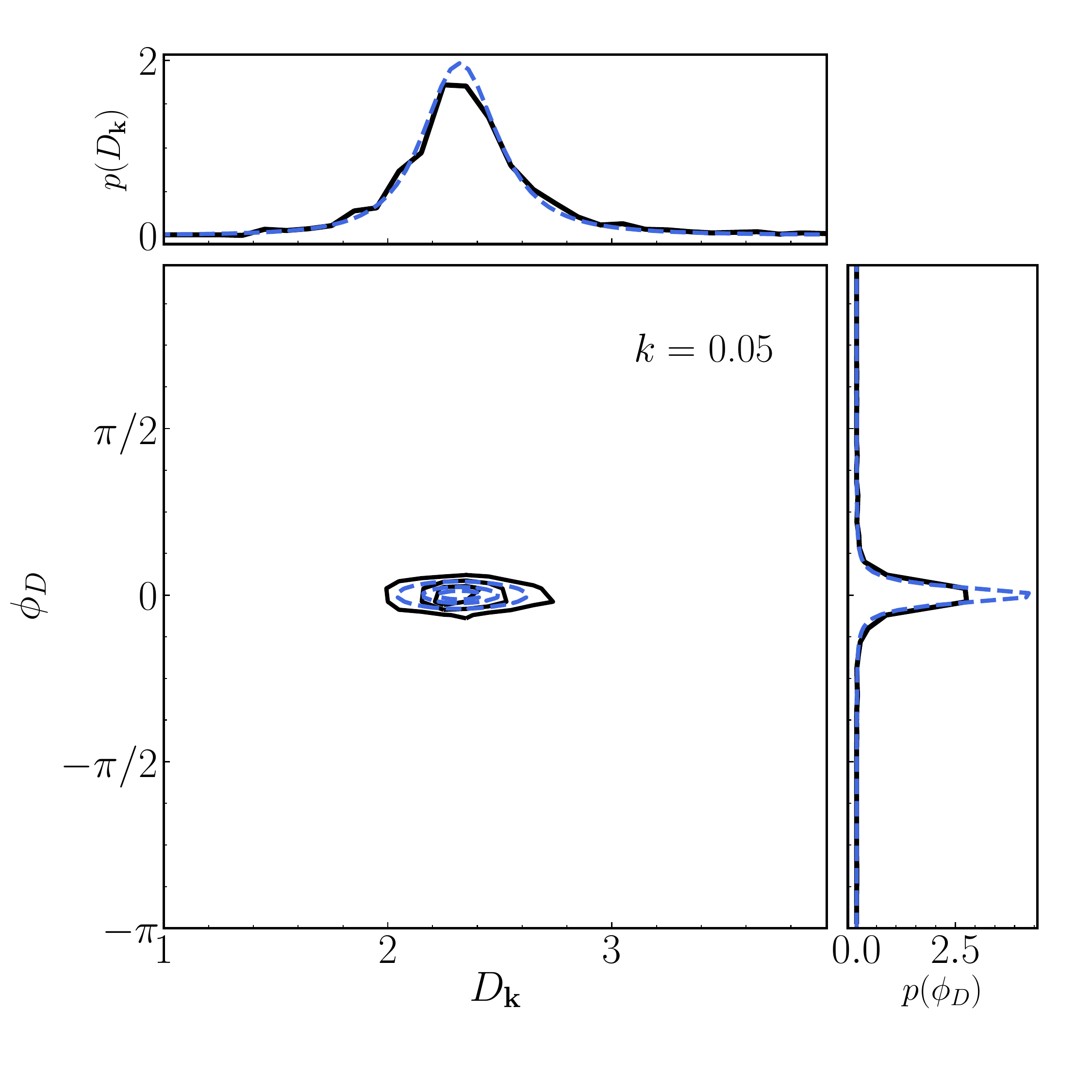}
\includegraphics{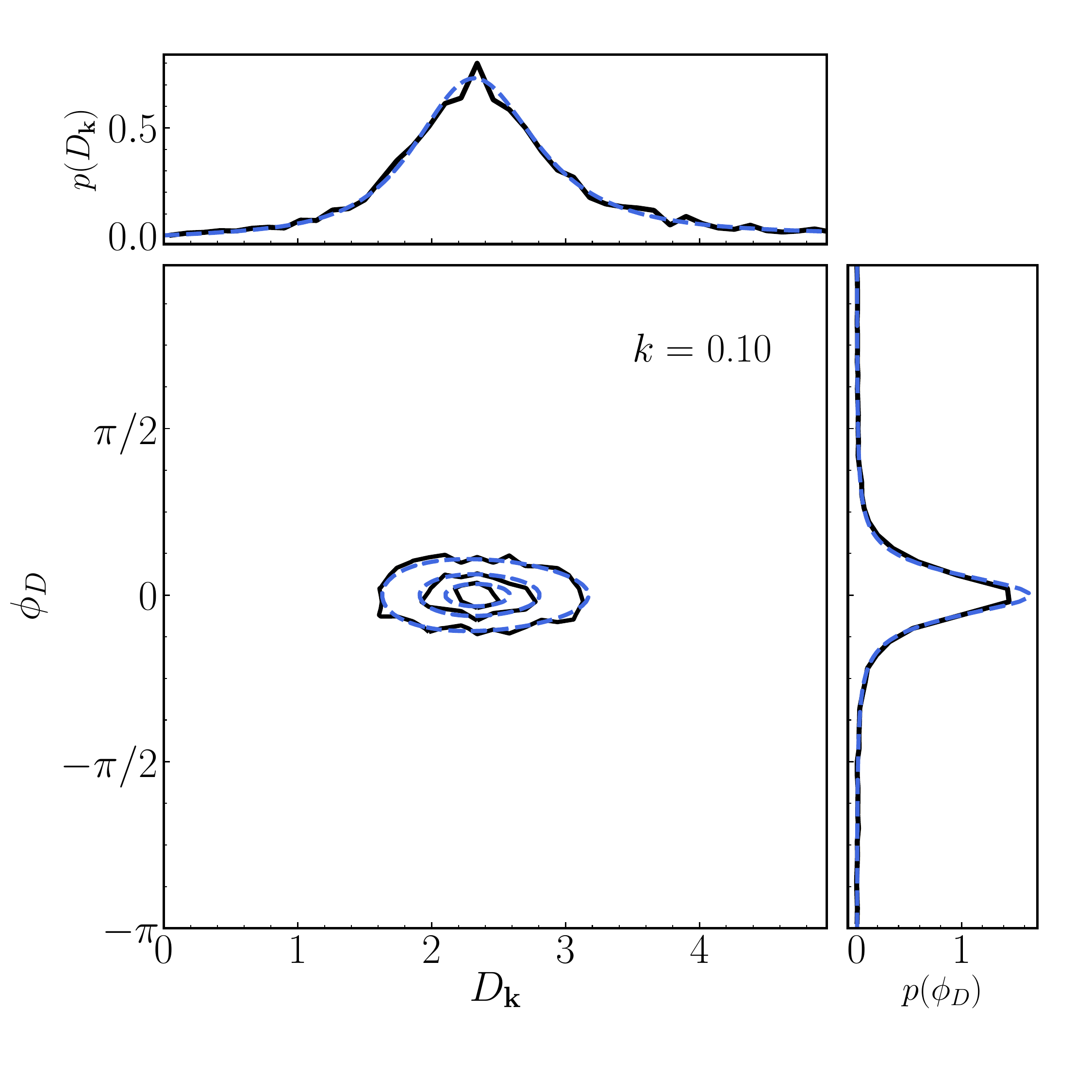}
\includegraphics{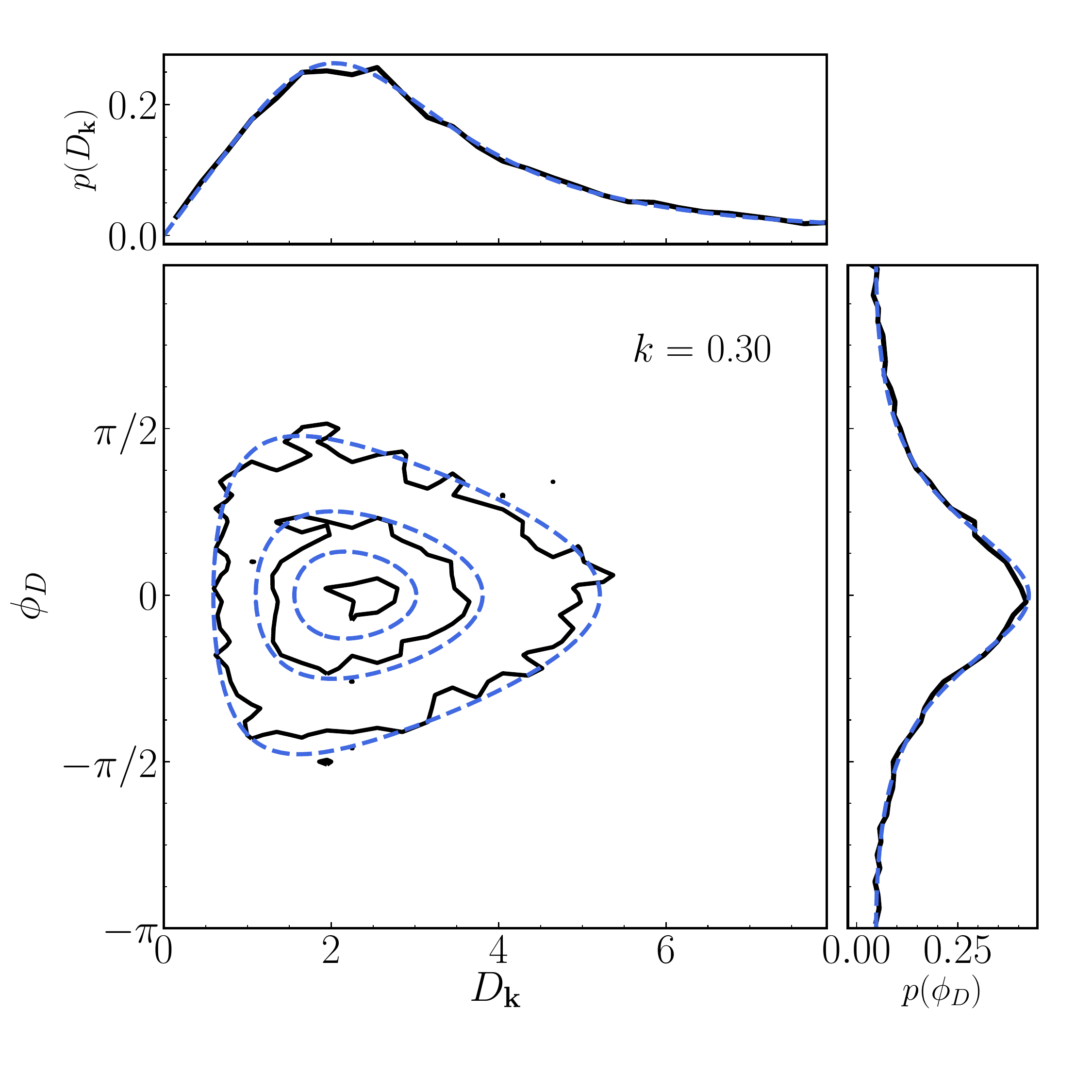}}


\caption{$p(D_\kvec, \phi_D)$ in the six $k$ bins as specified in Table \ref{tab:kf_bins}. The mode-dependent growth function ${\mathcal D}_\kvec=D_\kvec e^{i\phi_D}$ in this plot is of $z_1=0$ and $z_2=2$. In each panel, contour lines mark the levels at $20\%$, $50\%$ and $80\%$ of the maximum value of the distribution, $p(D_\kvec, \phi_D)$ is then integrated along each dimension to produce $p(D_\kvec)$ and $p(\phi_D)$ which are drawn in attached side subplots. The simulation results are colored with black solid lines, and the analytical models (Equation~(\ref{eq:Dk_pdf})) are plotted in blue dashed lines. $r=P_{z_1,z_2}/\sqrt{P_{z_1}P_{z_2}}$ and  $t=\sqrt{P_{z_1}/P_{z_2}}$ as inputs to the models are estimated from the simulation.}
\label{fig:delta_jpdf}
\end{figure*}

The one-point PDFs of ${\mathcal D}_\kvec$, $D_\kvec$ and $\phi_D$ in the selected $k$ bins are demonstrated in Figure \ref{fig:delta_jpdf}. The plot is generated by comparing the $z=0$ snapshot to the one at $z=2$ of the {\tt Pangu} simulation, together with the theoretical predictions of Equation~(\ref{eq:Dk_pdf}). The contours of $p(D_\kvec, \phi_D)$ measured from simulation are noisy, but $p(D_\kvec)$ and $p(\phi_D)$ appear to be highly consistent with the predictions of the model in all scale bins.

For large-scale modes ($k=0.05\ \ihmpc$), widths of the distributions $p(D_\kvec)$ and $p(\phi_D)$ are apparently narrow, indicating that the deviation from linear evolution is mild. With increasing $k$, the distributions become much wider, $p(\phi_D)$ tends to become uniformly distributed, and $p(D_\kvec)$ becomes skewed with a long tail. It implies that with the information provided by one-point statistics of the density field alone, one can hardly recover the cosmic density field exactly at earlier times in the nonlinear regime.

\subsection{Distribution of mode growth rate}
\label{sec:sim_zpdf}

\begin{figure*}[ht!]
\resizebox{\hsize}{!}{
\includegraphics{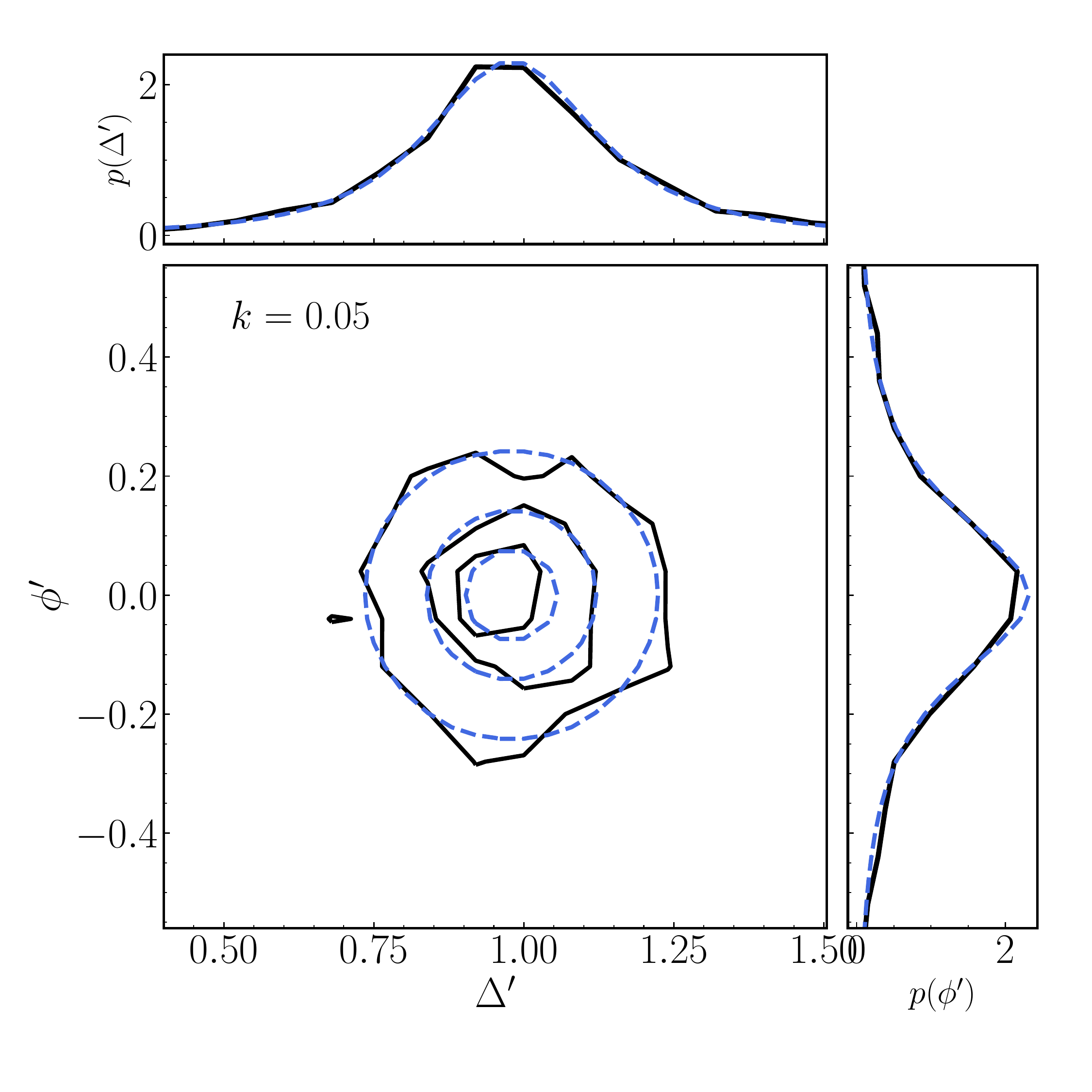}
\includegraphics{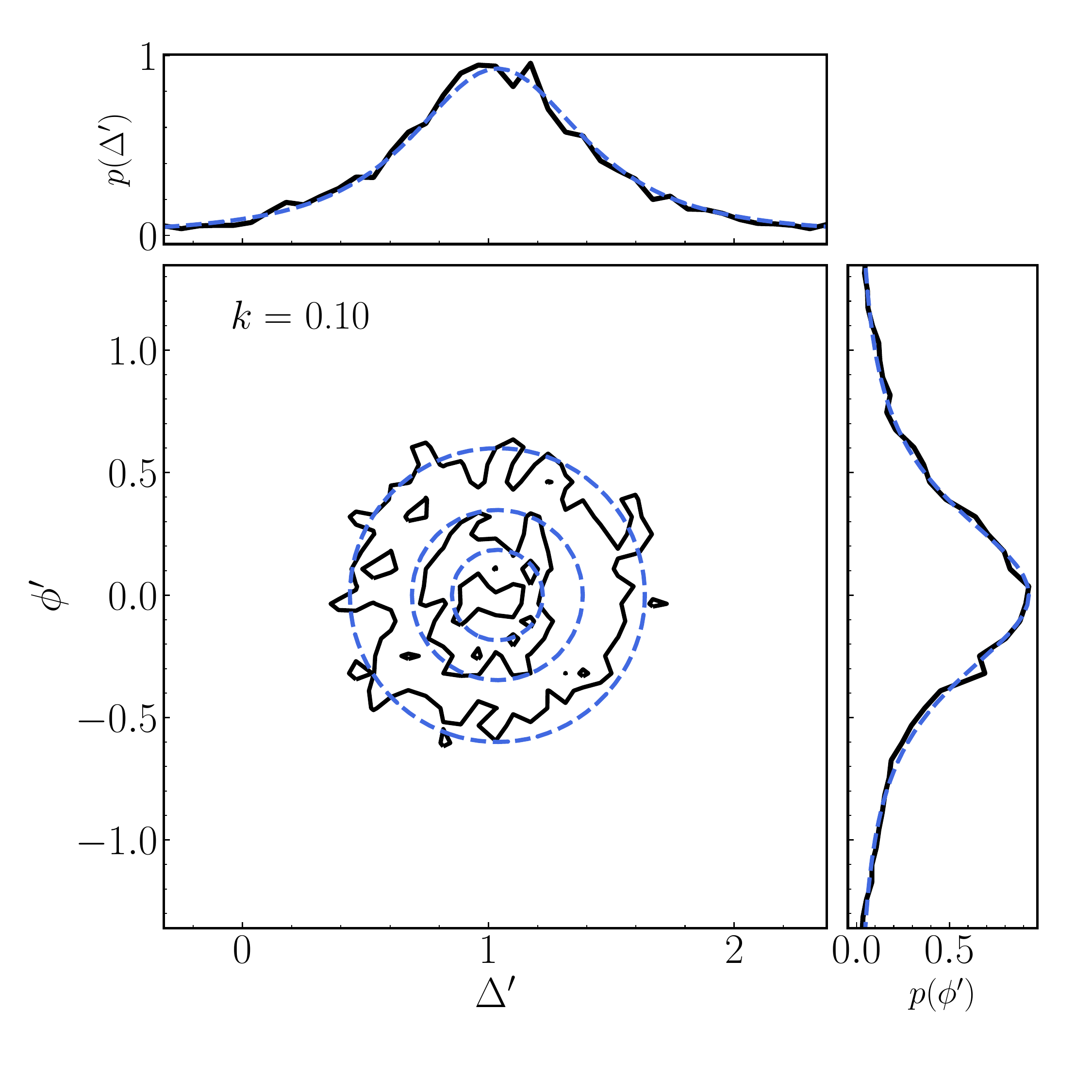}
\includegraphics{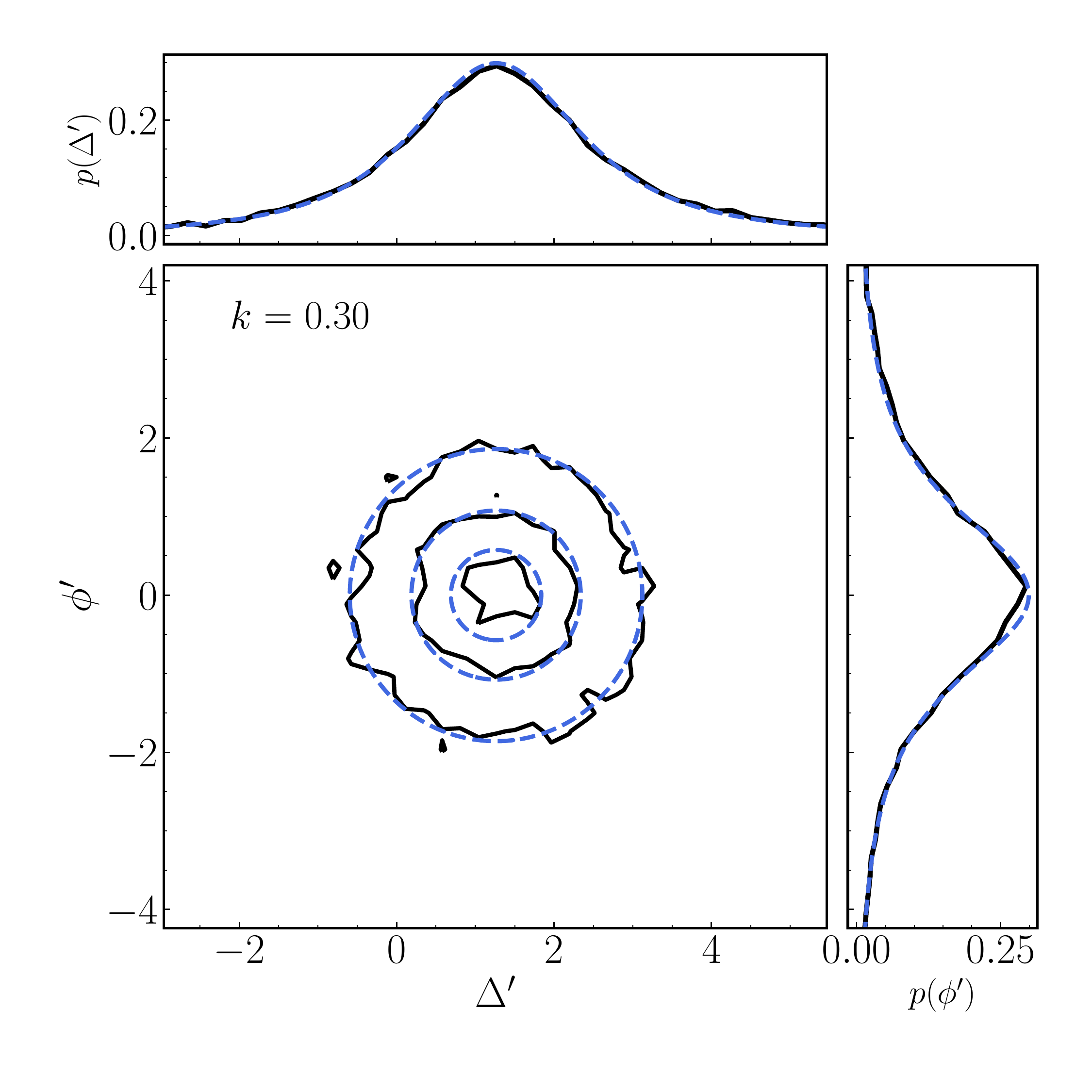}}

\resizebox{\hsize}{!}{
\includegraphics{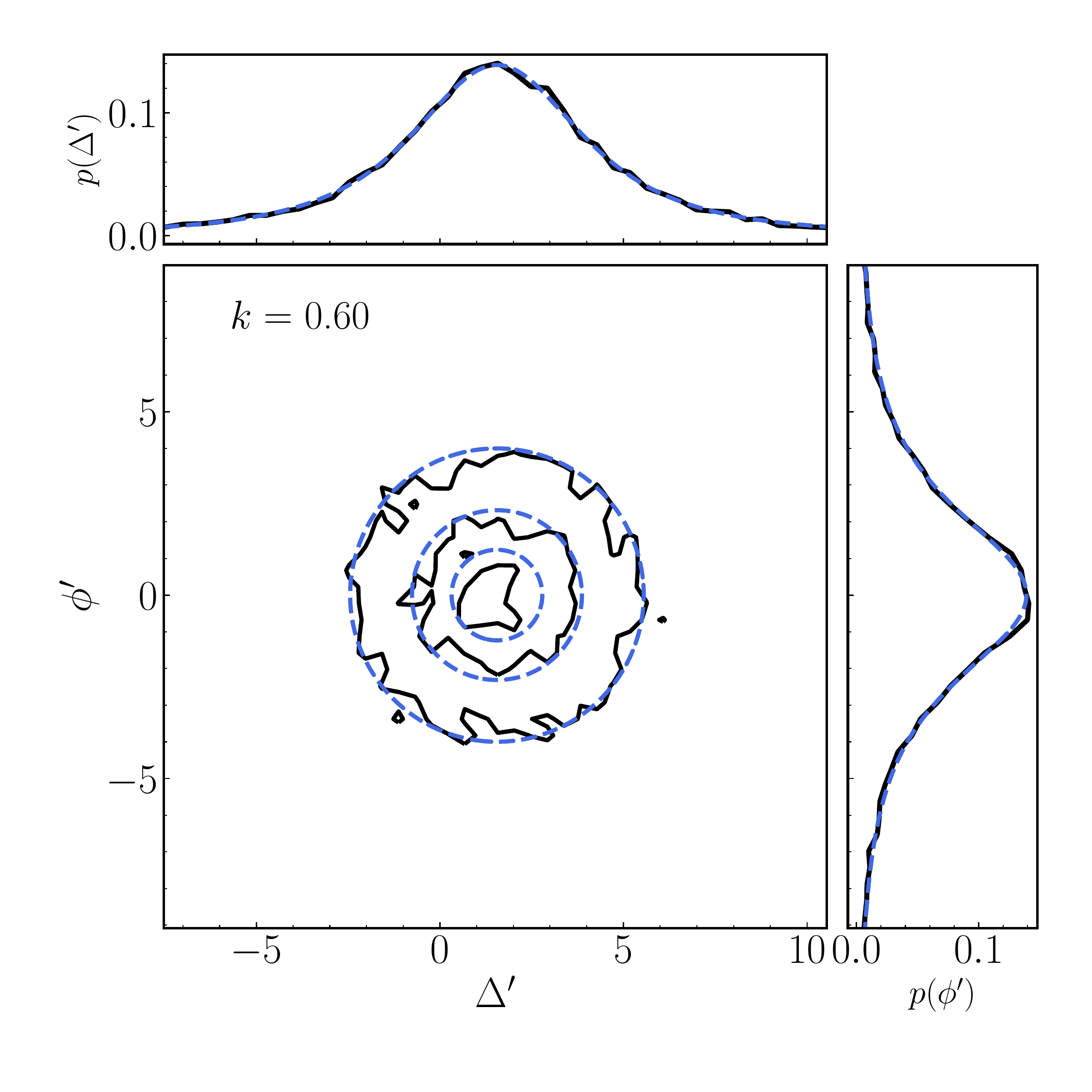}
\includegraphics{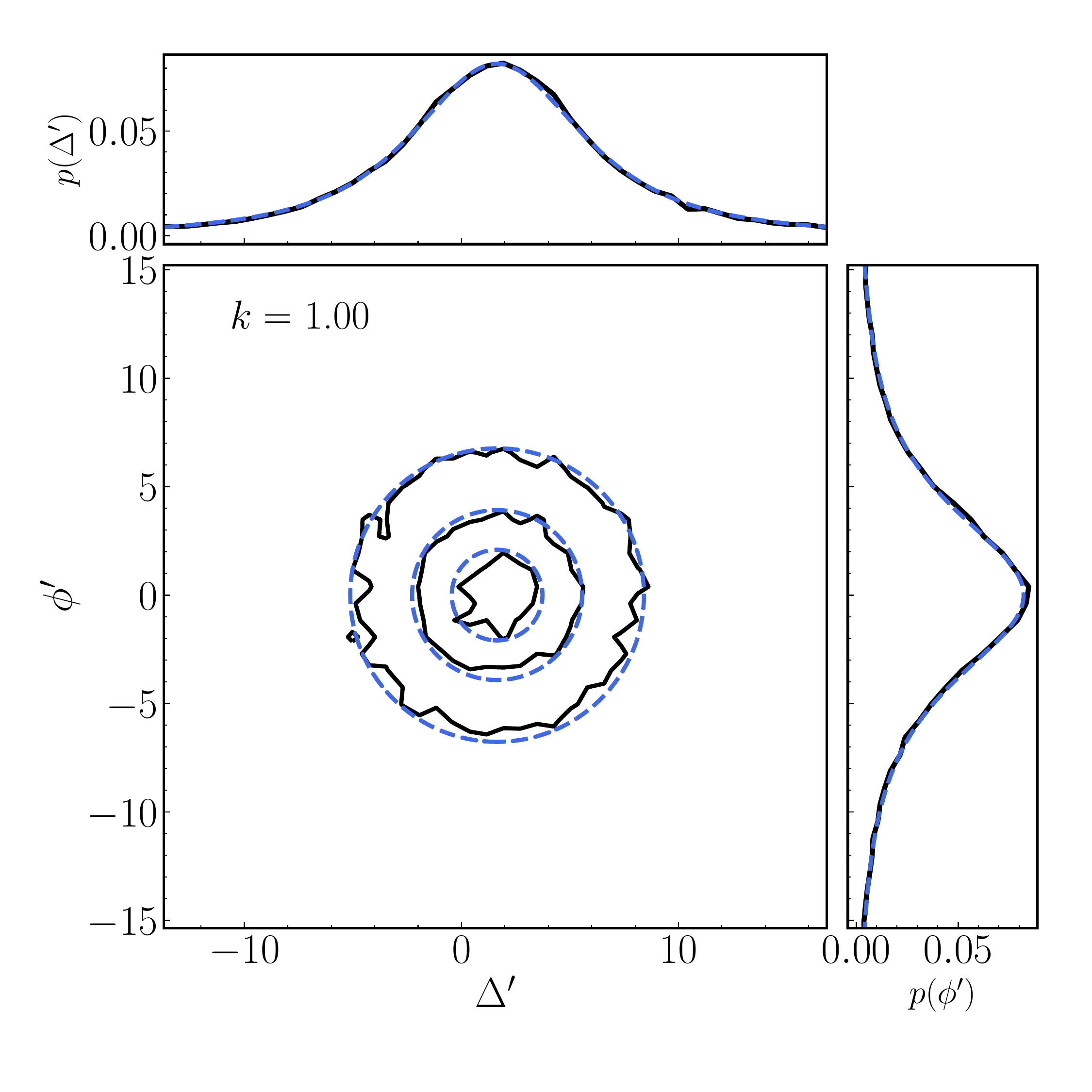}
\includegraphics{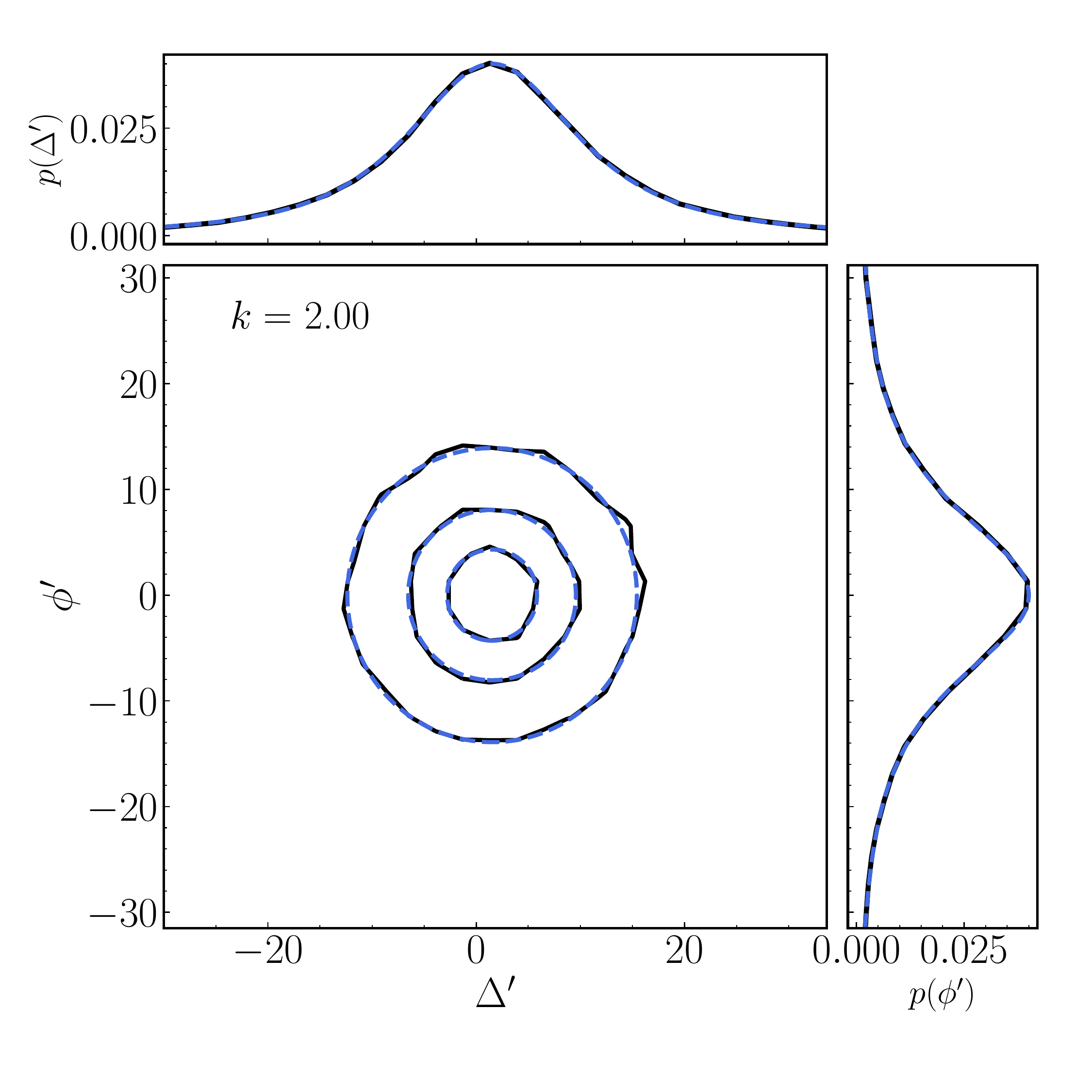}}

\caption{Similar to Figure \ref{fig:delta_jpdf}, but for $p(\Delta', \phi')$. $p(\Delta')$ and $p(\phi')$,  produced by integrating $p(\Delta', \phi')$, are shown in the side subplots attached in each panel. Analytical models (Equation~(\ref{eq:z_pdf})) are plotted in blue dashed lines. Again, $\avg{\Delta'}$, $\avg{\phi'}$, $\alpha$ and $\beta$ as input parameters to the models are taken from the simulation results.}
\label{fig:jpdf}
\end{figure*}

Figure~\ref{fig:jpdf} shows the one-point PDFs of $X_\kvec=\Delta'+i\phi'$ in selected $k$ bins calculated from the simulation data, together with the predictions of Equation~(\ref{eq:z_pdf}). The measurements can be well approximated by the model based on the Gaussian assumption. 

Since $\Delta'$ and $\phi'$ are following the Student's t-distribution with 2 degrees of freedom, their second-order moments do not exist, and the standard deviations numerically computed from the simulation data are meaningless. By Equation~(\ref{eq:z_pdf}), we propose taking an alternative quantity, $S\equiv \beta\sqrt{1-\alpha^2}$, as an appropriate function to characterize the width of $p(\Delta',\phi')$. As shown in Figure~\ref{fig:z_width} without surprise, the widths of $p(\Delta', \phi')$ become larger with increasing nonlinearity (from $z=2$ to $z=0$) . 
\begin{figure}
\plotone{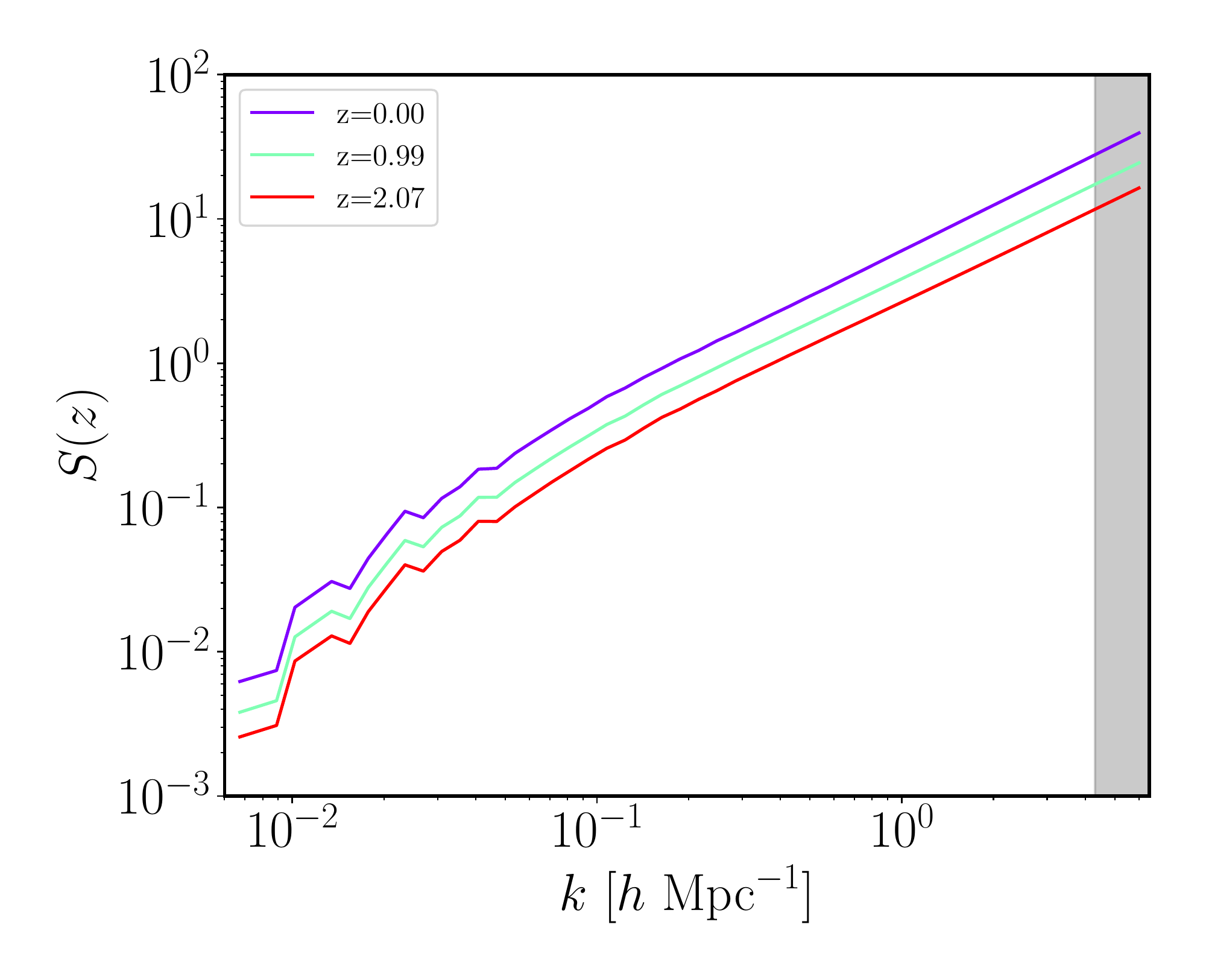}
\caption{Characteristic width parameter $S=\sqrt{\beta^2(1-\alpha^2)}$ of the one-point distribution $p(\Delta',\phi')$ as a function of scales.}
\label{fig:z_width}
\end{figure}
If we further decompose the momentum divergence into two independent components, $\psi=\psi_s+\psi_\delta$, with $\psi_s$ the stochastic part uncorrelated with density at all, and $\psi_\delta$ the part fully correlated with the density field that can be expressed as $\psi_\delta(\kvec)=T_{\delta \rightarrow \psi}(\kvec) \delta(\kvec)$ with a certain transfer function $T_{\delta\rightarrow \psi}$ in Fourier space. If one denotes the power spectrum of $\psi_s$ as $P_{\psi_s}$, it leads to 
\begin{equation}
S^2=(P_\delta P_\psi -P_{\delta\psi}^2)/P_\delta^2=P_{\psi_s}/P_\delta\ .
\end{equation}
From this decomposition, it turns out that the distribution widths of $p(\Delta', \phi')$ are mainly caused by the stochastic part of the momentum divergence $\psi_s$.

It is interesting to note that there is a crude scaling relation $S(z) \propto D(z)$ whose accuracy is better than $10\%$ within wide ranges of scale and redshift (Figure~\ref{fig:S_scaling}). A deviation greater than $5\%$ occurs mainly on scales $0.1 \lesssim k \lesssim 1 \ \ihmpc$ for $z>1$. By the one-loop Eulerian perturbation theory, $S$ scales with $D$ in the weakly nonlinear regime \citep[e.g.][]{SmithEtal2009, Pan2020}. But in the nonlinear regime, where the one-loop theory breaks seriously, the scaling relation still holds valid with considerable precision, and the performance turns out to be even better at lower redshifts, which is really intriguing. However, it is necessary to be cautious when quoting the precision of the scaling relation for $k< 0.1\ \ihmpc$, $S(z)$ is sensitive to the treatment of shot noise in the weakly nonlinear regime (see Appendix~\ref{sec:app2} for more details).
 
\begin{figure}
\plotone{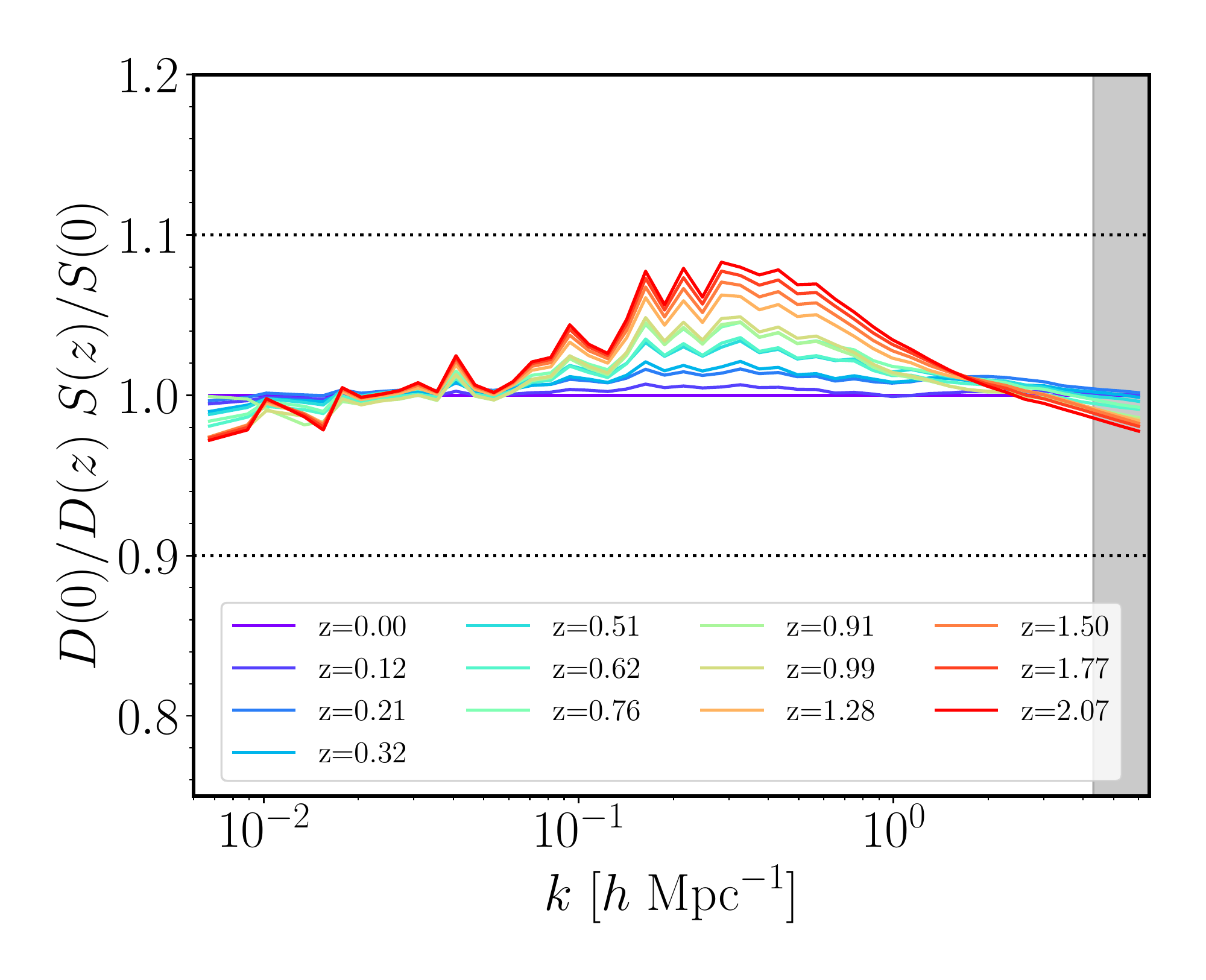}
\caption{$[S(z)/D(z)]/[S(0)/D(0)]$ measured at multiple redshifts $0 \leq z \lesssim 2$. The shaded regime is where $k>0.67\ k_N$. Here, the shot noise in the power spectra used to compute $S$ is not subtracted.}
\label{fig:S_scaling}
\end{figure}

Since the symmetry of $\left[ \delta(-\kvec), \psi(-\kvec) \right]=\left[ \delta^*(\kvec), \psi^*(\kvec) \right]$ has ensured $\avg{\phi'}=0$,  $\avg{\Delta'}$ is then of major interest. As already mentioned in Equation~(\ref{eq:z_mean}), $\avg{ \psi_\kvec \deltak^*/\deltak\deltak^*}$ is equal to $\avg{\psi_\kvec \deltak^*}/\avg{\deltak\deltak^*}$. This means that $P_{\delta\psi}/P_\delta$ is a good estimator of $\avg{\Delta'}$. In Figure~\ref{fig:mean_error}, we demonstrate the effectiveness of Equation~(\ref{eq:z_mean}) with simulation data. The mean differences between $P_{\delta\psi}/P_\delta$ and $\avg{\psi_\kvec/\deltak}$ are generally less than $1\%$ within the scales $0.007 \lesssim k \lesssim 6\ \ihmpc$. 

\begin{figure}
\plotone{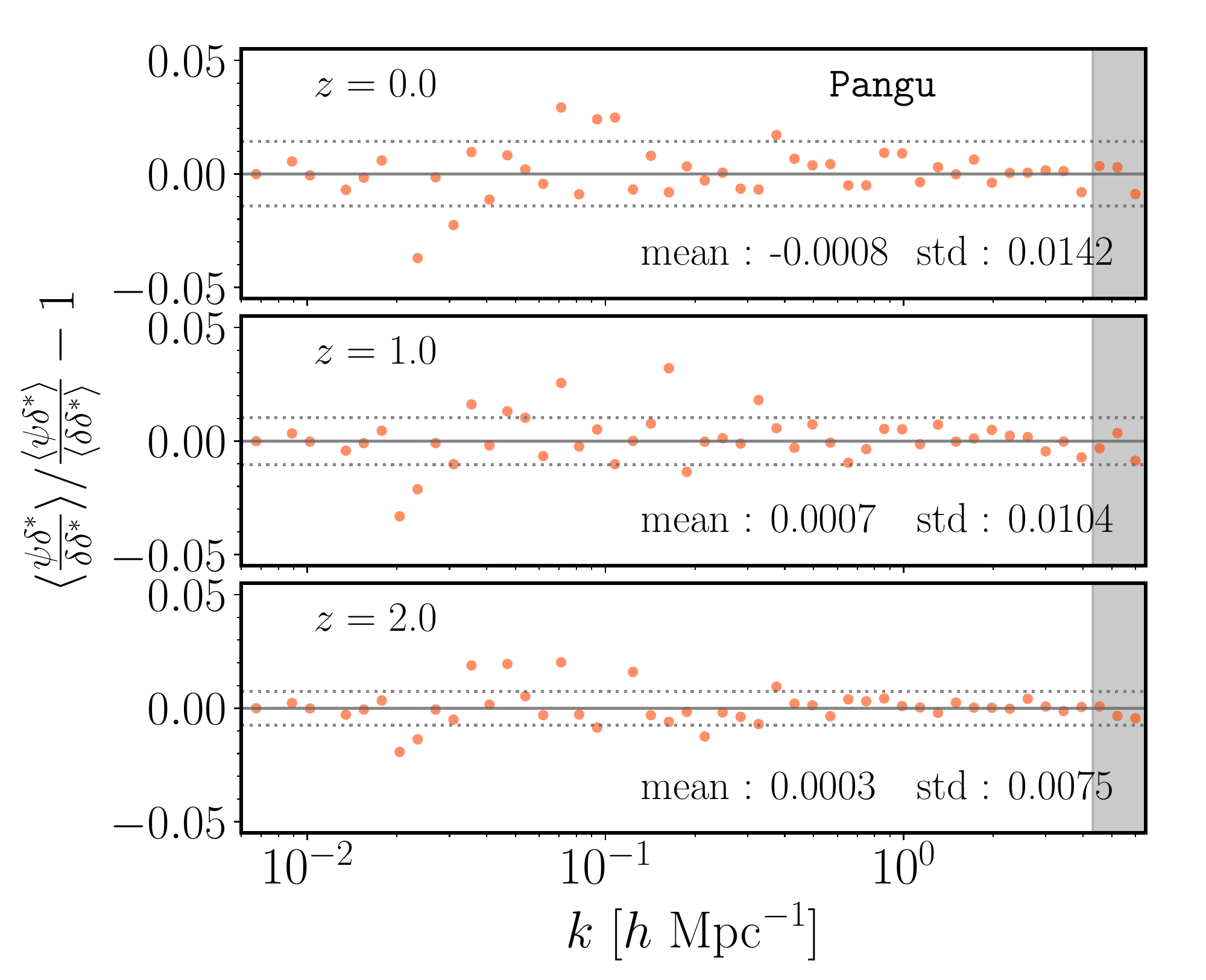}
\caption{Relative differences between $P_{\delta\psi}/P_\delta$ and $\avg{\psi_{\kvec}/\deltak}$. Data from three snapshots at $z=0,1,2$ of the {\tt Pangu} simulation are used. Horizontal solid lines represent the mean of the differences, and dashed lines delimit the standard deviations. In each subplot, the means and standard deviations are calculated from all data points along with $k$ scales. The gray shaded area marks the regime where $k>0.67\ k_N$.}
\label{fig:mean_error}
\end{figure}

\subsection{The mean mode growth rate in simulation}
\label{sec:z_mean_simu}
The mean mode growth rate $\avg{\Delta'}$ is measured from the simulation data using $P_{\delta\psi}/P_\delta$. Its dependence on multiple redshifts and scales is fully investigated and shown in Figure~\ref{fig:zmean_sim}. On large scales $k \lesssim 0.1\ \ihmpc$, $\avg{\Delta'}\approx 1$ at all redshifts, which is consistent with the linear theory for gravitational evolution. In the intermediate nonlinear regime, $\avg{\Delta'}$ rises with increasing $k$, and is systematically greater when the redshift approaches zero. The value of $\avg{\Delta'}$ has a single peak positioned between $1 \sim 4\ \ihmpc$, and the peak height is less than 2. The general trend is that the peaks of $\avg{\Delta'}$ at higher redshifts are located at larger $k$ than those at lower redshifts, but the relations among peak position, peak height, and redshift are not simple.

\begin{figure*}
\plottwo{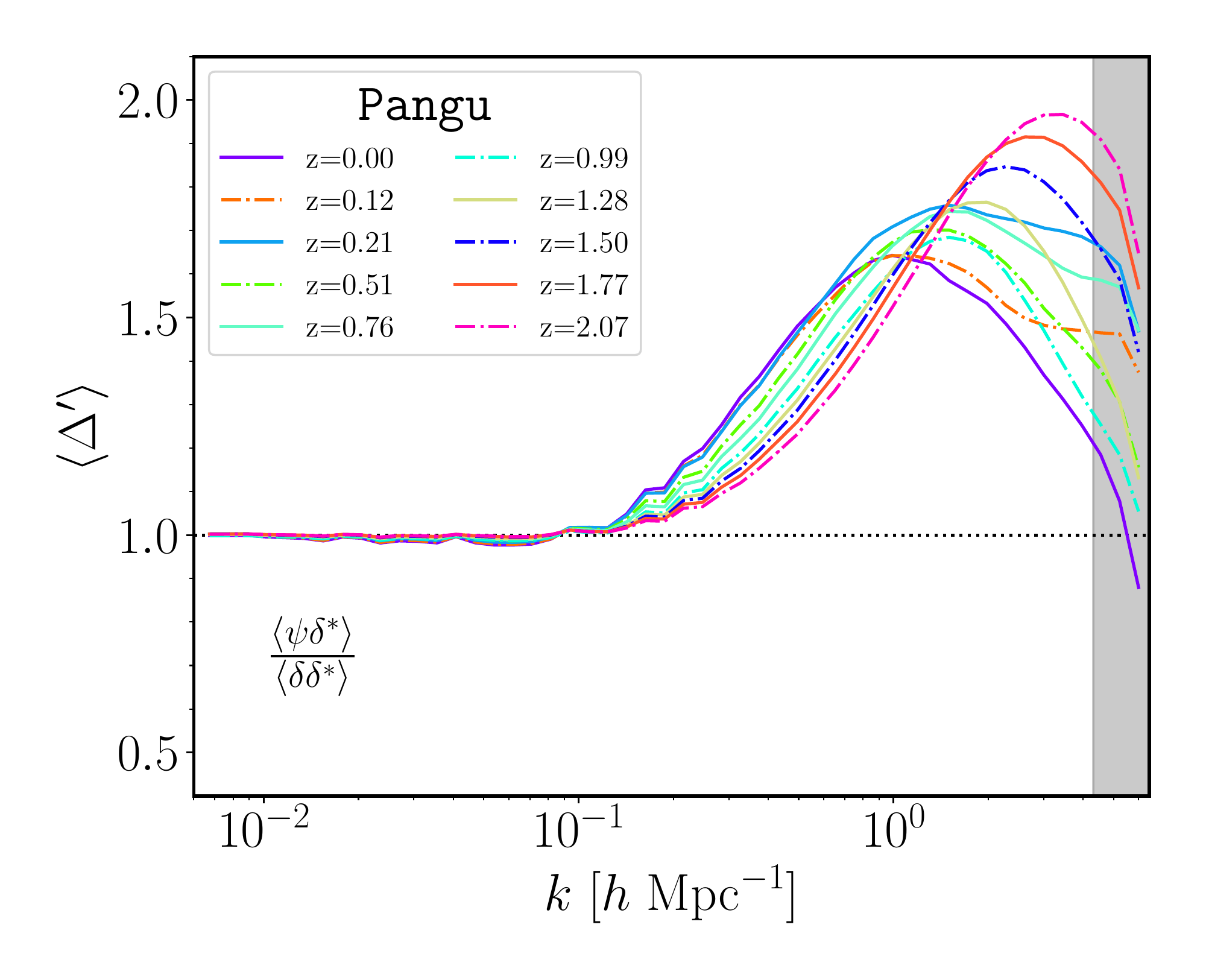}{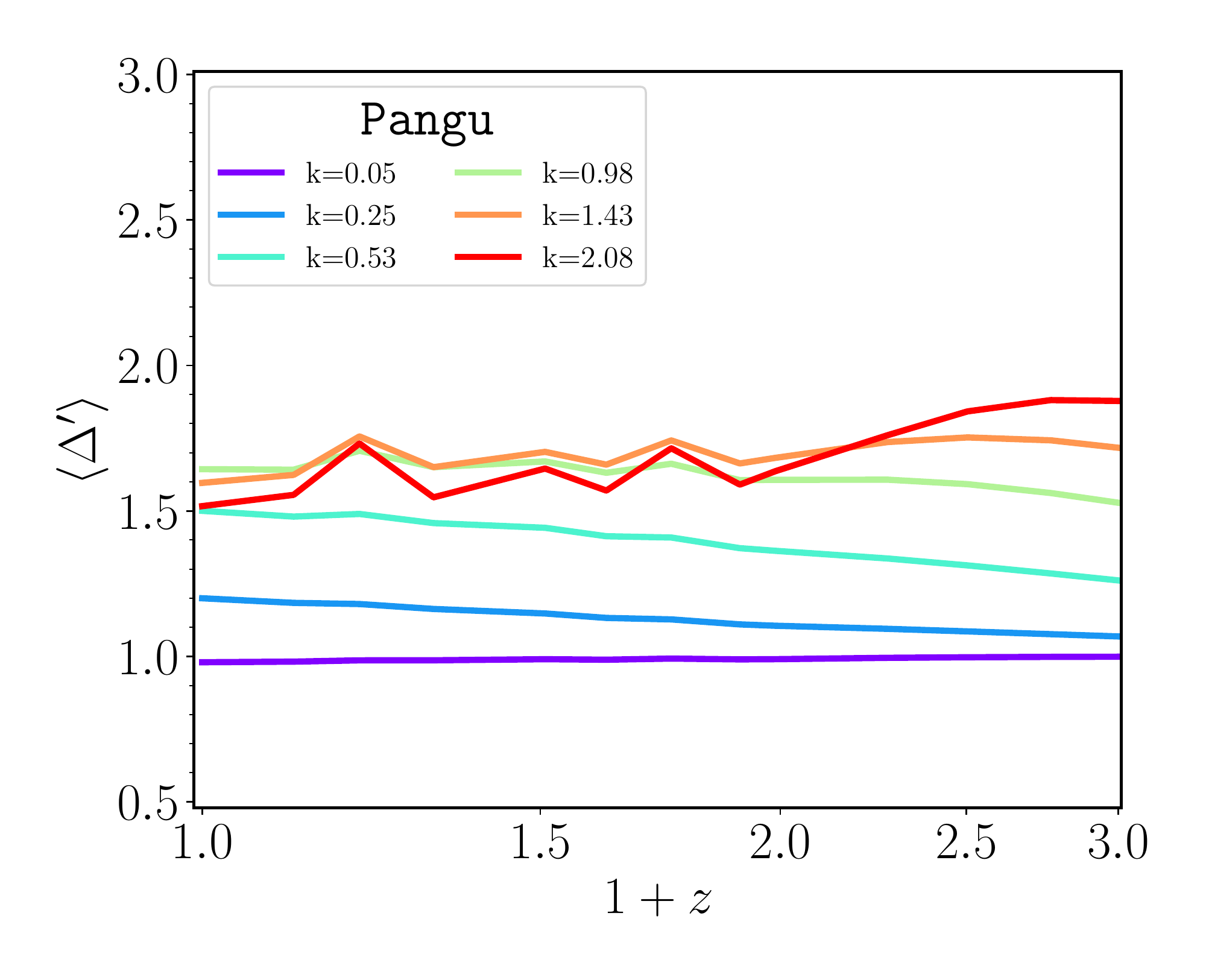}
\caption{Mean mode growth rates $\avg{\Delta'}$ estimated from the simulation by $P_{\delta\psi}/P_\delta$, as functions of scales (left panel) and redshifts (right panel). Lines of various colors label $\avg{\Delta'}$ of different redshifts (left panel) and scales (right panel) accordingly. Again, the gray region in the left panel is of $k>0.67\ k_N$.}
\label{fig:zmean_sim}
\end{figure*}

The scale dependence of $\avg{\Delta'}$ is shown in the right panel of Figure~\ref{fig:zmean_sim}. In general, the amplitude of $\avg{\Delta'}$ along the redshift is stronger for higher $k$, but $\avg{\Delta'}$ at $k \gtrsim 1\ \ihmpc$ tends to decrease when $z\rightarrow 0$, contrary to its behavior at $k\lesssim 1\ \ihmpc$. In addition, non-trivial oscillatory structures are also found at $k \gtrsim 1\ \ihmpc$. The origins of these phenomena are still unclear. The measurement is based on only one simulation realization and could be camouflaged by cosmic variance\footnote{Please see Appendix~\ref{sec:app3} to get a rough idea of the influence of cosmic variance on $\avg{\Delta'}$.}. One potential solution is to have a large number of simulation realizations with $\sim h^{-3}{\rm Gpc}^3$ volume and mass resolution better than $\sim 10^9\ \msun$. This is a prerequisite for relevant investigation toward smaller scales, probably to $k\sim 10\ \ihmpc$. We leave this for further work.

\section{Empirical models for the mean mode growth rate in nonlinear regime}
\label{sec:zmean_model}
\subsection{Fitting formulae}
\label{sec:fitting}
Developing theoretical models of $\avg{\Delta'}$ in the strongly nonlinear regime from first principles is a difficult task, desperately in demand of ingenious ideas. Therefore, we turn to building templates for nonlinear $\avg{\Delta'}$ empirically in different cosmologies. The solution is already provided by Equation~(\ref{eq:zmean_pk}), together the algorithm based on the first-order finite difference method. We first pick up a fitting formula for $P_\delta$ and construct a table as a function of redshift $z$, with a fine increment of $\Delta z$. Then $\avg{\Delta'}$ at a given redshift $z$ is calculated numerically with the derivative
\begin{equation}
    \frac{\ln P_\delta(z-\Delta z)-\ln P_\delta(z+\Delta z)}{2\Delta \ln D(z)}, 
\end{equation}
where $\Delta \ln D(z) = \ln D(z-\Delta z)- \ln D(z+\Delta z)$, $z \pm \Delta z$ are two interpolation boundaries that enclose $z$ in the constructed $P_\delta$ table. We have tried different values of $\Delta z$ and $\Delta \ln D$, and once the result becomes stable as $\Delta \ln D\rightarrow 0$, we take it as a phenomenological model of $\avg{\Delta'}$.

In this work, emulators of {\tt HMCODE 2020} \citep{MeadEtal2021},  {\tt EUCLID EMU} \citep[version 2;][]{KnabenhansEtal2020}, {\tt BACCO EMU} \citep{AnguloEtal2020}, {\tt MIRA TITAN} \citep{LawrenceEtal2017},  and the empirical model of {\tt Takahashi 2012} \citep{TakahashiEtal2012}, are chosen to generate the nonlinear power spectrum. In Figure~\ref{fig:model2sim}, the performance of these prescriptions is presented in comparison to the simulation results. These fitting formulae do recover the power spectrum very well, but cannot accurately reproduce $\avg{\Delta'}$ in the strongly nonlinear regime, for example, $k\gtrsim 1\ \ihmpc$ for all shown cases. Again, the results presented are of a single simulation set and more realizations are needed to draw reliable conclusions. Nevertheless, one thing is sure that it seems $\avg{\Delta'}$ can be a useful tool for distinguishing the performance of different fitting formulae of the power spectrum in the strongly nonlinear regime.

\begin{figure*}
\plottwo{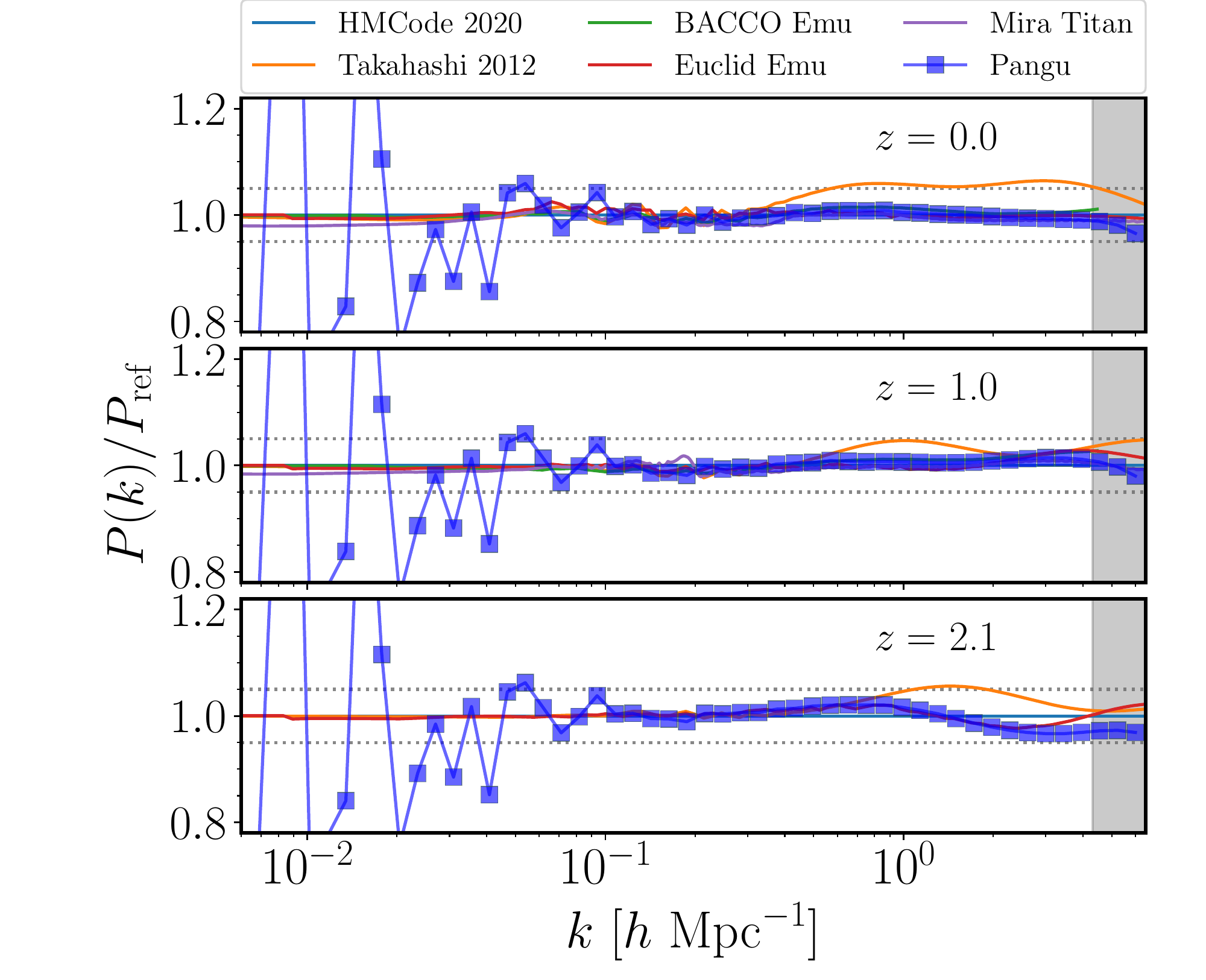}{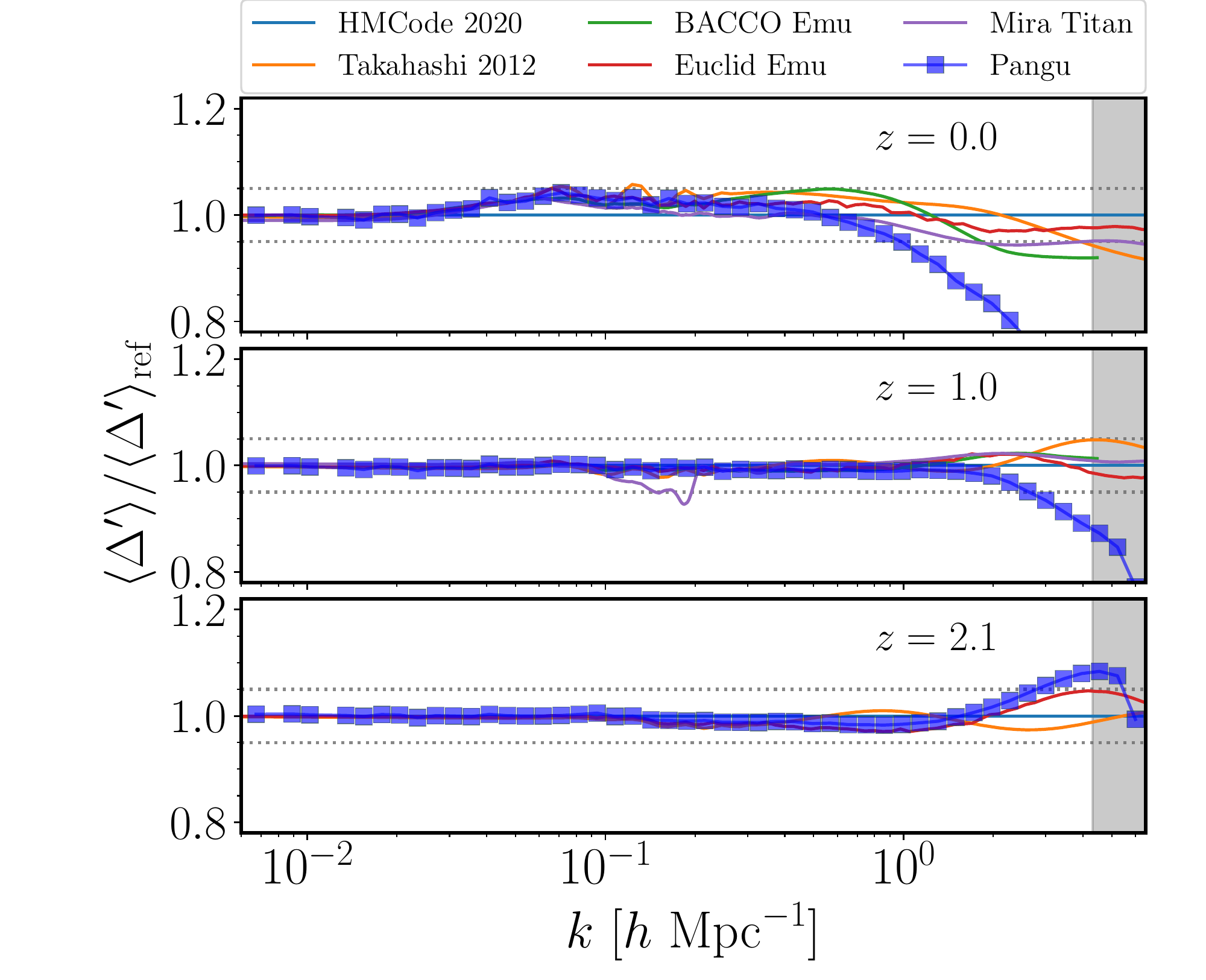}	
\caption{Comparison between simulations and empirical models, in terms of power spectrum (left panel) and $\avg{\Delta'}$ (right panel). Solid lines of different colors are given by fitting formulae as specified on the top of each panel, square points connected with blue lines are measurements of simulations, and dashed horizontal lines are used to mark the $\pm 5\%$ differences. The shaded regions are where $k>0.67\ k_N$. The reference template (quantity with subscript {\rm "ref"}) is given by {\tt HMCODE 2020} \citep{MeadEtal2021}. }
\label{fig:model2sim}
\end{figure*}

\subsection{Scale transformation}
\label{sec:scale_trans}
It is well known that the nonlinear dimensionless power spectrum of the dark matter density field, $\Delta^2(k)=P(k)k^3/(2\pi^2)$, can be approximated as a function of the linear dimensionless power spectrum $\Delta^2_L(k_L,z)=(D(z)/D(0))^2 \Delta^2_L(k_L, 0)$ after a scale transformation, 
\begin{equation}
k_L=[1+\Delta^2(k)]^{-1/3}  k.
\label{eq:k_scaling}
\end{equation}
The scaling {\em ansatz} was first proposed to model the two-point correlation function by \citet{HKLM1991}, then developed in the Fourier space by \citet{PeacockDodds1994}. It is found that Equation~(\ref{eq:k_scaling}) could also be applied to approximately recover the bispectrum when nonlinearity is not strong \citep{PanEtal2007}. Along with Equation~(\ref{eq:zmean_pk}), we conjecture that the transformation of Equation~(\ref{eq:k_scaling}) could help to understand $\avg{\Delta'}$. 

\
\begin{figure*}
\resizebox{\hsize}{!}{
\includegraphics{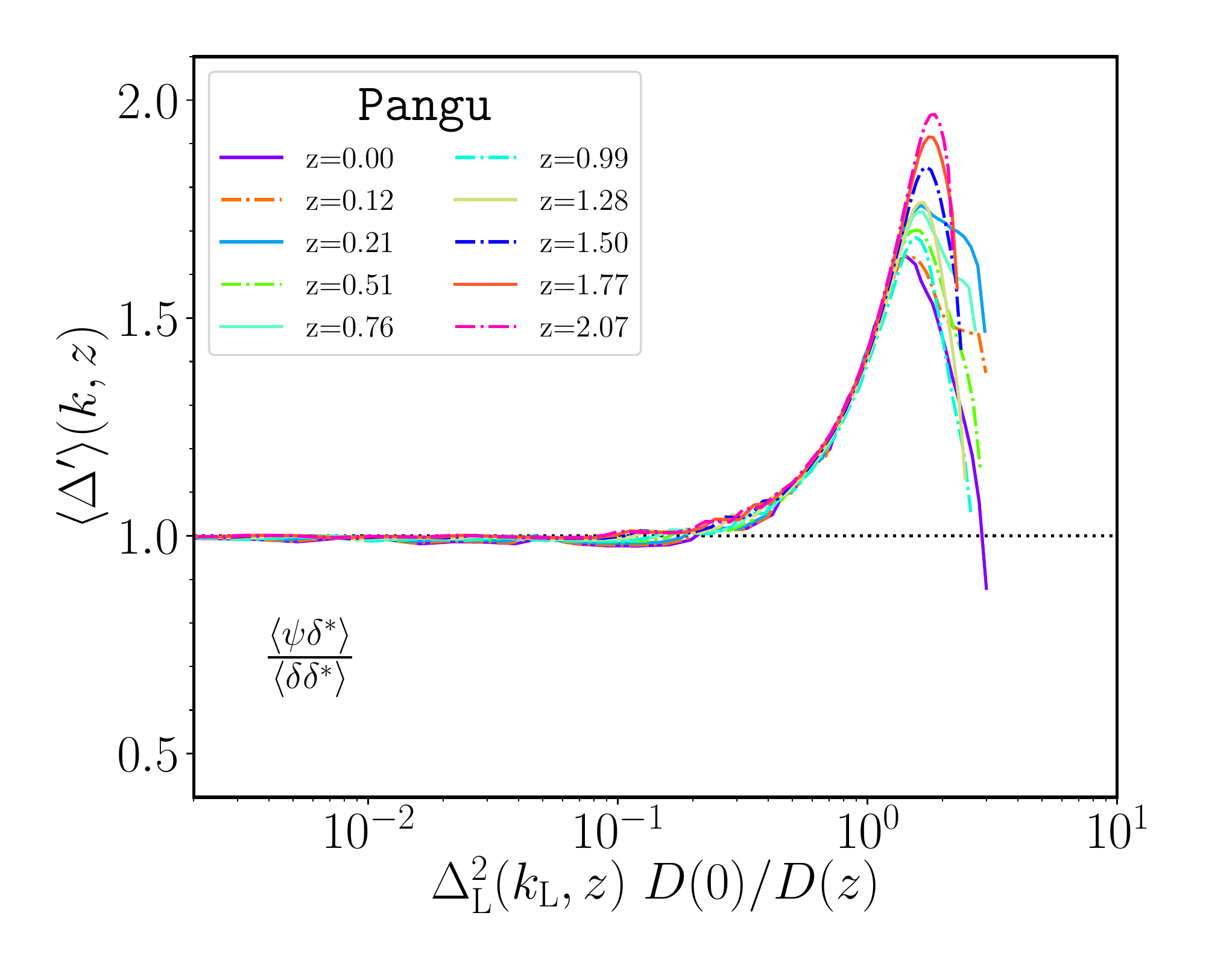}
\includegraphics{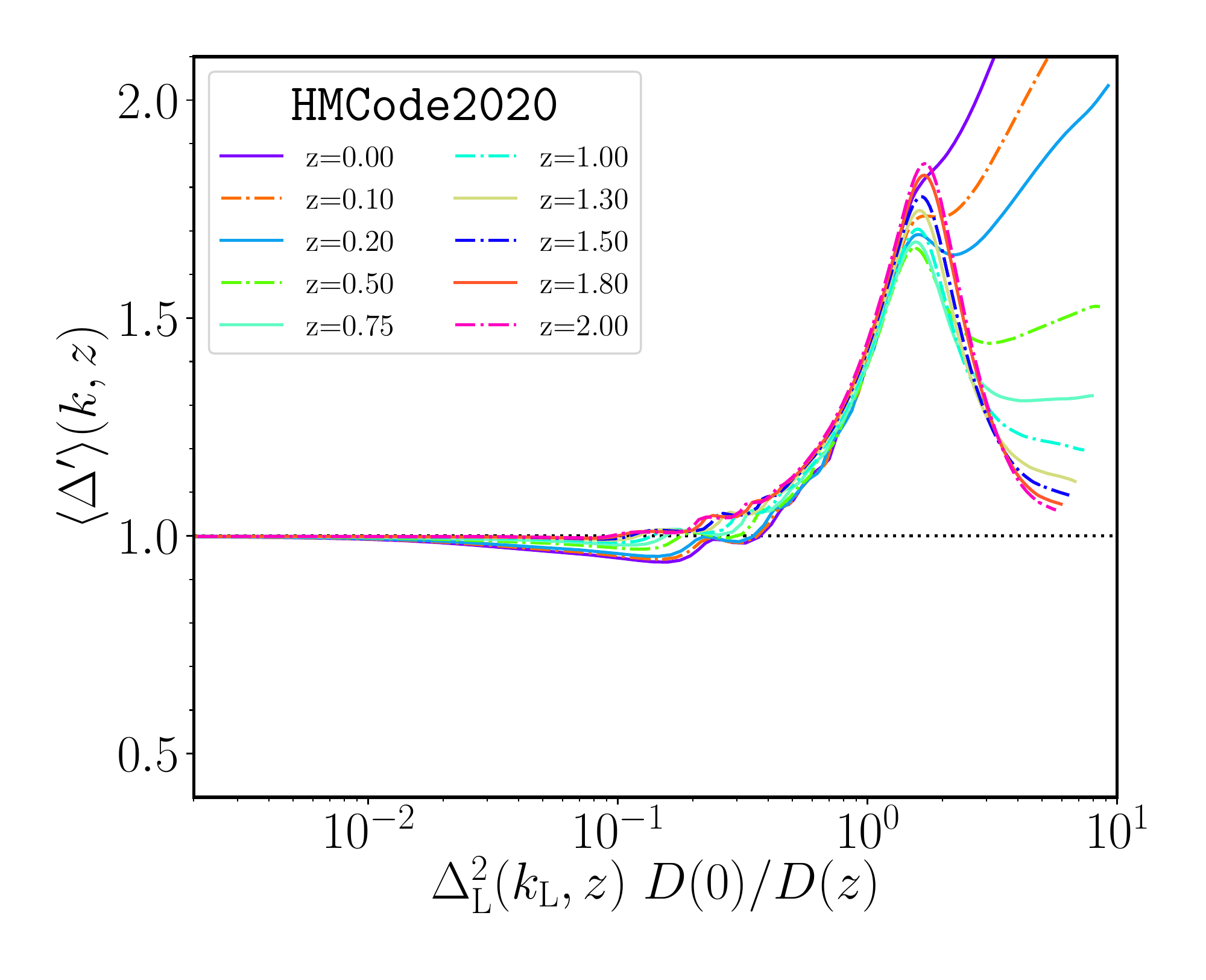}
\includegraphics{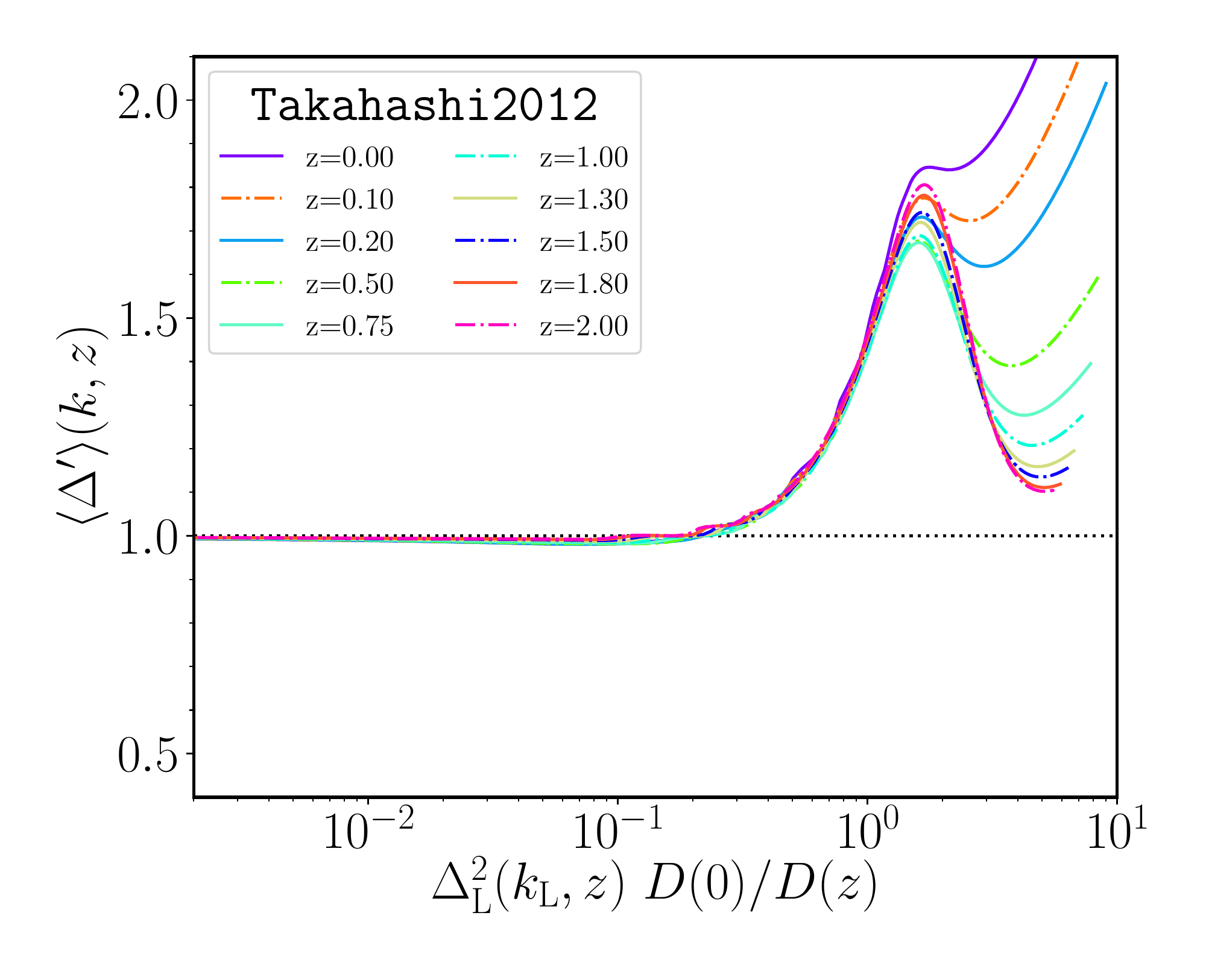}}
\caption{The mean mode growth rates as functions of $\Delta_L^2(k_L, z)/(D(z)/D(0))$, $k_L$ are given by Equation~(\ref{eq:k_scaling}). The left panel shows the results computed from the {\tt Pangu} simulation, the middle and right panels are the results of the emulator {\tt HMCODE 2020} \citep{MeadEtal2021} and the empirical model {\tt Takahashi 2012} \citep[][]{TakahashiEtal2012} accordingly with the same cosmology.}
\label{fig:zmean_scaling}
\end{figure*}
\

The two formulae of {\tt HMCode 2020} and {\tt Takahashi 2012} are called to facilitate our experiment. The results of the simulation {\tt Pangu} simulation and the two models are shown in Figure~\ref{fig:zmean_scaling}. It turns out that the scale transformation of Equation~(\ref{eq:k_scaling}) is really useful, although an extra factor is needed in practice. For $\Delta^2_L D(0)/D(z) \lesssim 1.3$, the mean mode growth rates at different epochs can be written as
\begin{equation}
\avg{\Delta'(k,z)}= F[\Delta^2_L(k_L, z)D(0)/D(z)]\ ,
\label{eq:zmean_scaling}
\end{equation}
rather than $F[\Delta^2_L]$ as one might naively expect. Here, $F[\ldots]$ refers to a general function of a certain form that may contain additional dependencies on the cosmological parameters, and the accuracy is better than $5\%$.

The peaks of $\avg{\Delta'}$ of the simulation and the two empirical models are basically relocated to a narrow range of $\Delta^2_L D(0)/D(z) \sim 1.3-2$ after the scale transformation. But the differences between the simulation and the empirical models are still very prominent for $\Delta^2_L D(0)/D(z) \gtrsim 1.3$. Empirical models of {\tt HMCode 2020} and {\tt Takahashi 2012} (the middle and right panels of Figure~\ref{fig:zmean_scaling}) produce rising tails on scales greater than peak locations, and these tails arise systematically higher at lower redshifts, which are not present in the ${\tt Pangu}$ simulation. Considering that empirical models are obtained through extensive calibration against simulations, the evolution of dark matter clustering in very strongly nonlinear regimes may be sensitive to numerical details of different simulations. One needs to be careful when using fitting formulae as templates if their $\Delta^2_L D(0)/D(z)$ goes beyond 1.3.

\section{Summary}
\label{sec:summary}
In this study, we demonstrate that nonlinear cosmic fields such as the dark matter density $\delta$ and the momentum divergence $\psi$ can be effectively and efficiently Gaussianized by the Fourier transformation at the one-point level, which is an extension of the work of \citet{Matsubara2007}. Gaussianity of the one-point distributions of Fourier modes greatly simplifies analysis of the spatial distribution of dark matter and its evolution, as what has been shown in this work about functions given by the ratio of two complex random variables.

Analytical formulae about the one-point PDF of the quotient of two correlated complex Gaussian random variables are introduced, and then applied to explore statistical properties of two quantities used to describe the clustering evolution of dark matter. The first is the complex mode-dependent growth function $\mathcal{D}_\kvec(z_1,z_2)$, defined as the ratio of $\deltak$ at two different epochs, $X_\kvec = \delta_\kvec(z_1)/\delta_\kvec(z_2)$. The explicit model of the one-point PDF of $\mathcal{D}_\kvec$ is in good agreement with the results of the N-body simulation. The amplitude of the complex mode-dependent growth function $D_{\kvec}$ and its logarithm are also investigated. It is really intriguing that, for the dark matter density, the ratio of the cross-power spectrum to the auto-power spectrum and the ratio of two auto-power spectra have particular statistical meanings. The former one is the mean of $\mathcal{D}_\kvec$, while the square root of the latter is the median of the amplitude $D_{\kvec}$.

Another instance studied in this work is the complex mode growth rate, which is defined in Fourier space by the ratio $X_\kvec=\psi_\kvec/\deltak=\Delta'+i\phi'$. With the continuity equation, we identify that the real part of $X_\kvec$, $\Delta'$ is the growth rate of the amplitude of the density mode, while the imaginary part $\phi'$ relates to the phase growth rate. The Gaussian approximation to the one-point PDF $p(\Delta', \phi')$ is again developed and confirmed by simulation with good accuracy. It turns out that the mode's amplitude growth rate $\Delta'$ and the mode's phase growth rate $\phi'$ both follow the Student's t-distribution with 2 degrees of freedom. The distribution could be characterized by an alternative width parameter $S$, which increases with the strength of nonlinearity, and has an approximate scaling relation $S \propto D$ among different epochs. 

As $\avg{\phi'}$ is always zero, the information about the nonlinear evolution process of the density field in $X_\kvec$ is mainly packed in the mean mode growth rate $\avg{\Delta'}$. On large scales where nonlinearity is very weak, $\avg{\Delta'}\approx 1$ at all redshifts. When goes to the nonlinear regime, the increasing nonlinearity will drive $\avg{\Delta'}$ away from unity. We further show that empirical formulae and theoretical models of the density power spectrum can provide satisfactory templates for $\avg{\Delta'}$ in the weakly and intermediate nonlinear regime, but the agreement begins to drop drastically in the strongly nonlinear regime. The phenomenon becomes even more apparent after applying a scale transformation.

In summary, with the Gaussian approximation to cosmic fields in Fourier space, the statistical properties about the quotient of two complex Gaussian variables provide a novel approach to measure the growth of individual mode of dark matter density field. It has been proven to be very useful and can be readily applied in other large-scale structure studies with similar types of quantities. The results presented here emphasize the simplicity of the analysis in Fourier space and are worthy of more attention for further studies of cosmic structure formations.

\begin{acknowledgements}
Ming Li acknowledges support from the NSFC grants of Nos. 11988101, 12033008. Jun Pan acknowledges support from the NSFC grants of No. 11573030. Pengjie Zhang acknowledges support from the NSFC grant No. 11621303. Longlong Feng acknowledges support from the NSFC grant No. 11733010. Guoliang Li thanks the support from NSFC grant No. U1931210. Weipeng Lin thanks the NFSC grant No. 12073089. Haihui Wang is supported by the Shandong MSTI Project (2019JZZY010122) and the MIIT grant (J2019-I-0001).

The authors thankfully acknowledge computing and storage support from the cosmology simulation database (CSD) in the National Basic Science Data Center (NBSDC) and its fund NBSDC-DB-10, No. 2020000088.

We would like to thank the anonymous referee for helping us improve this work with their constructive comments.
\end{acknowledgements}


\appendix

\section{Aliasing}
\label{app:a}
In the ratio of $P_{\delta\psi}/P_\delta$, the aliasing effect induced by the sampling function cannot be completely eliminated. We adopted the third-order Battle-Lemari\'{e} spline function to assign simulation particles onto FFT grids with different resolutions, $N_{\mathrm{grid}}=512, 1024, 2048$. Then the mean growth rates $\avg{\Delta'}$ are measured to carry on the convergence test. The results are shown in Figure \ref{fig:reso_test}. We find that a difference greater than $1\%$ occurs on the scales $k> \sim 0.7\ k_N$, where $k_N$ is the Nyquist frequency, which is consistent with \citet{Pan2020}. After careful examination, we determine that the maximal $k$ in this work is $0.67\ k_N$.

\begin{figure}
	\plotone{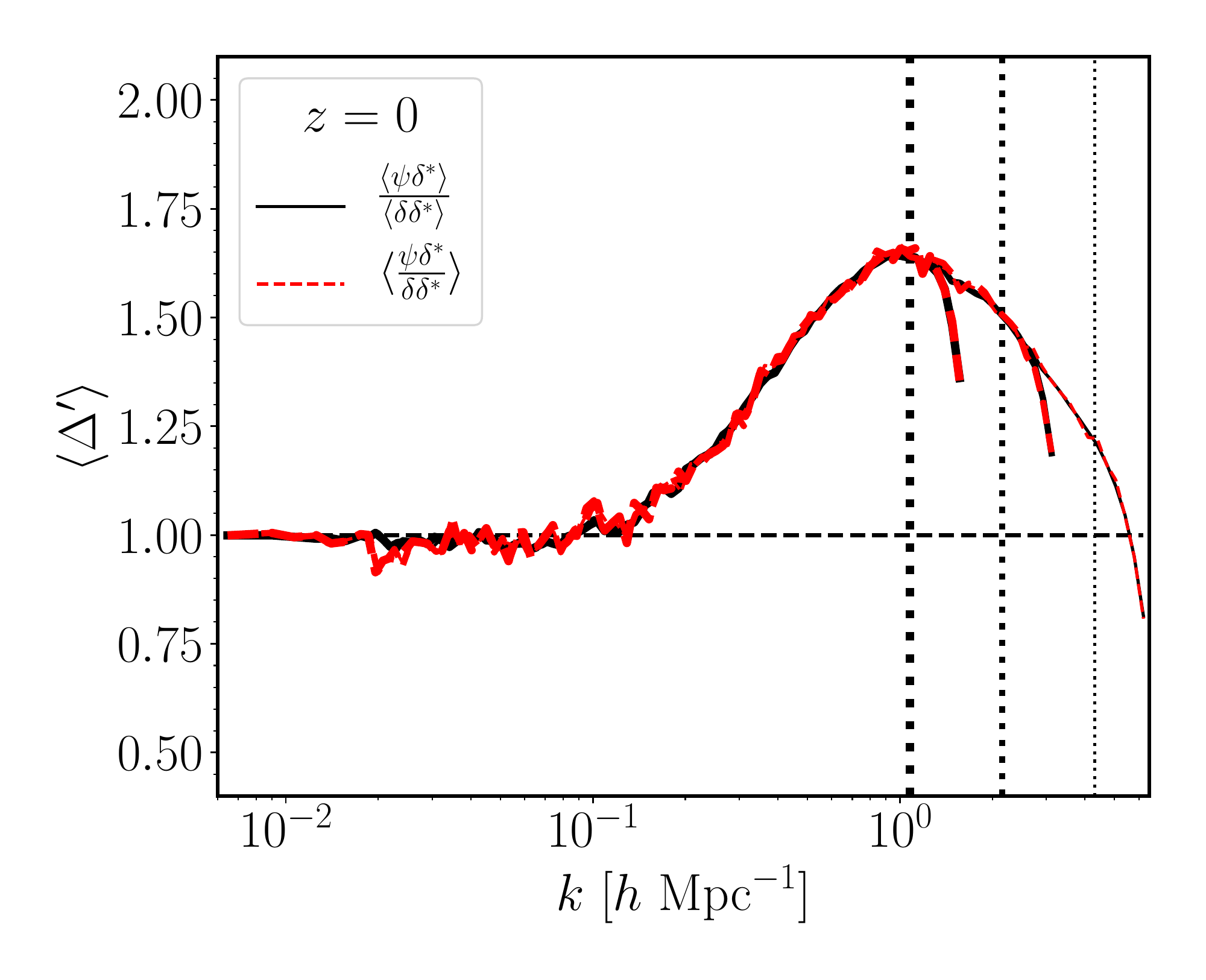}
    \caption{Measured mean mode growth rate with different grid numbers used for FFT. Vertical dotted lines delimit the scales of $k= 0.67\ k_N=1.08, 2.16, 4.31\ \ihmpc$ corresponding to grid number  $N_{\mathrm{grid}}=512, 1024, 2048$ respectively. Thinner curves are of finer resolution.}
    \label{fig:reso_test}
\end{figure}

\section{Effect of shot noise deduction on $S$}
\label{sec:app2}
The shot noises in $P_\delta$ and $P_\psi$ are minuscule, as the number of particles in our simulations is fairly large. As shown in Figure~\ref{fig:width_noise}, after subtracting the shot noise predicted by the local Poisson approximation \citep[e.g.][]{Pan2020}, the resulting $[S(z)/D(z)]/[S(0)/D(0)]$ does not obey $S\propto D$ well on scales $k<0.02\ \ihmpc$, compared to Figure~\ref{fig:S_scaling}. It is probably due to the fact that on large scales $P_\delta P_\psi -P^2_{\delta\psi} \approx 0$, a slight change will have a significant influence. But in the quasi-linear regime, the one-loop perturbation theory actually predicts $S\propto D$, whether this is a challenge to the way of subtracting shot noises is beyond the scope of this paper and will be left for future research.

\begin{figure}
\plotone{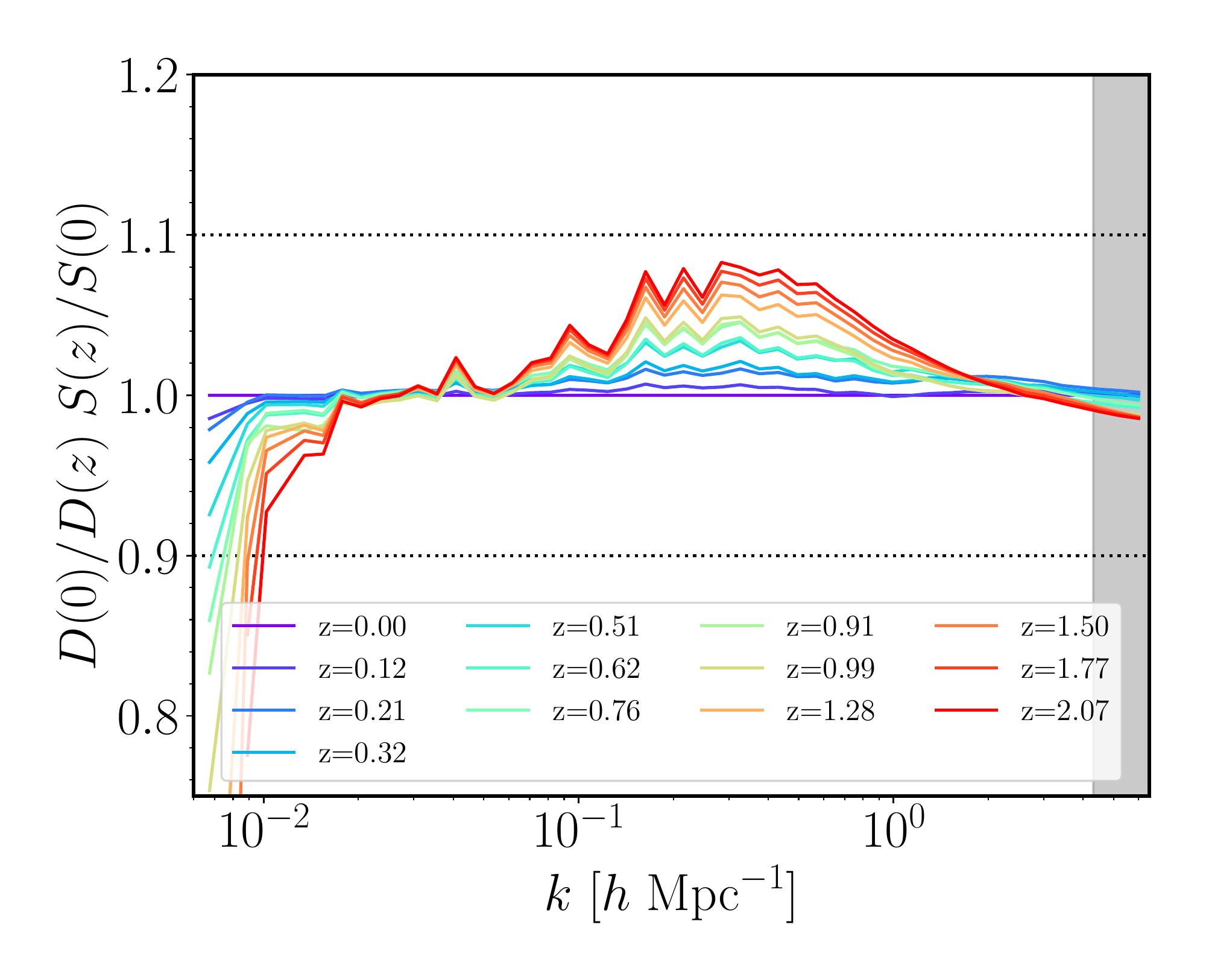}
\caption{Similar to Figure~\ref{fig:S_scaling}, but $[S(z)/D(z)] /$ $[S(0)/D(0)]$ of the {\tt Pangu} simulation is measured with Poisson noises subtracted from $P_\delta$ and $P_\psi$.} 
\label{fig:width_noise}
\end{figure}

\section{About cosmic variance}
\label{sec:app3}
We have only one realization of the simulation used in this report. To have a concept of the magnitude of cosmic variance in $\avg{\Delta'}$ on small scales, we split the $z=0$ snapshot of the {\tt Pangu} simulation into $8\times8\times8$ non-overlapping small subsamples of size $L_\mathrm{sub-box}=125\ \lhmpc$. The mean mode growth rates of these subsamples are measured and plotted together in Figure~\ref{fig:cv}. One can see that, although the 1-$\sigma$ scatter of $\avg{\Delta'}$ in the strongly nonlinear regime is not too large, there are indeed quite a few special subsamples whose $\avg{\Delta'}$ show quite different behaviors (brown solid lines in Figure~\ref{fig:cv}). 

\begin{figure}
\plotone{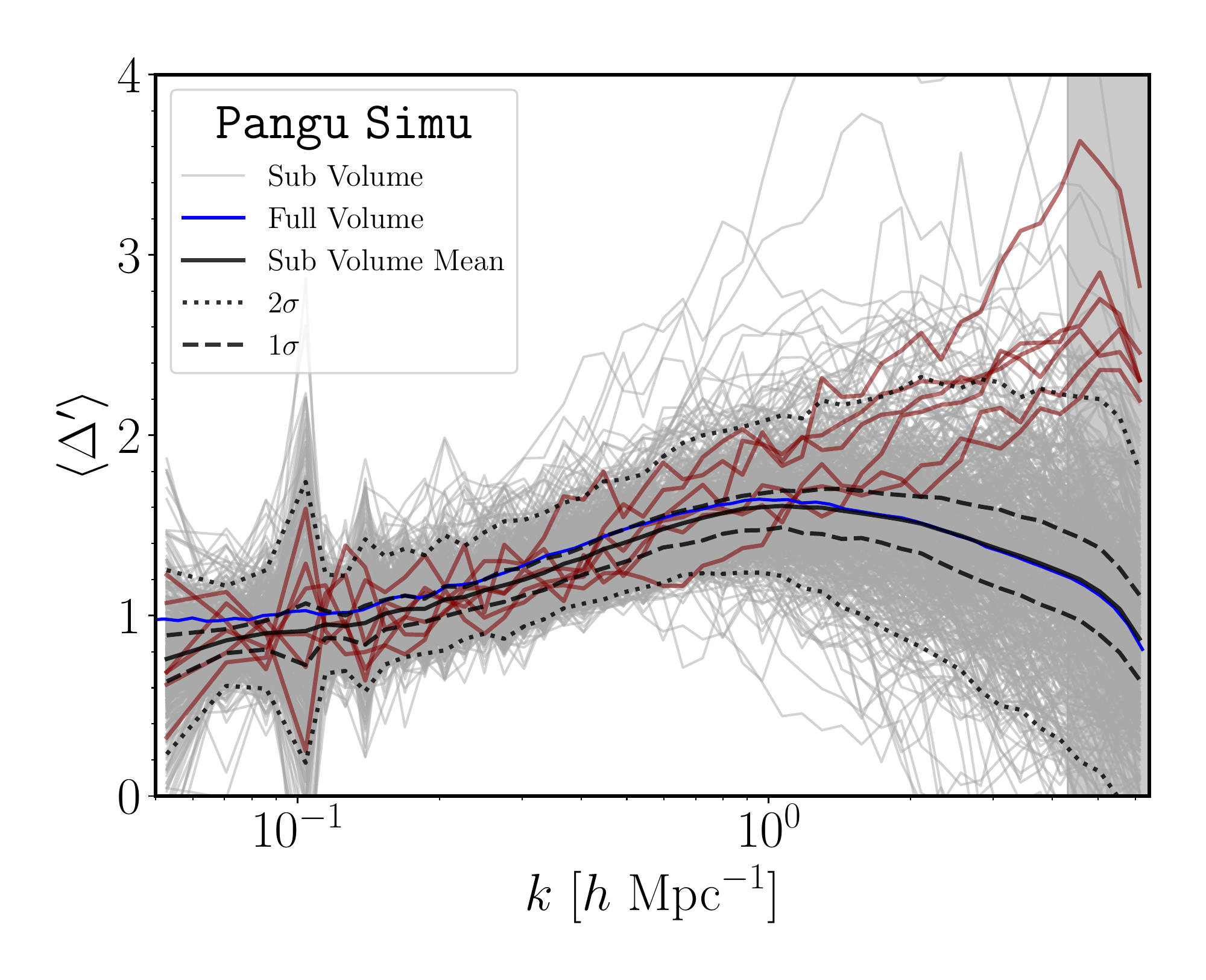}
\caption{Mean mode growth rates $\avg{\Delta'}$ measured from 512 subsamples, which are generated from the $z=0$ snapshot of the {\tt Pangu} simulation. Individual measurements are plotted as light gray solid lines, while their average and 1$\sigma$ and 2$\sigma$ scatters are drawn as black solid, dashed, and dotted lines, respectively. Brown solid lines are of the six selected subsamples that do not have apparent peaks on scales $k<0.67\ k_N$ at all, contrary to the measurement of the entire sample (solid blue line).} 
\label{fig:cv}
\end{figure}

\bibliographystyle{aasjournal}
\bibliography{vsmode_refs} 

\end{document}